\documentclass[preprint,aps,nofootinbib]{revtex4-1}

\usepackage{graphicx}

\usepackage{amsmath}
\usepackage{amsfonts}
\usepackage{amssymb}
\usepackage{color}
\usepackage{natbib}
\usepackage{hyperref}
\usepackage[USenglish]{babel}

\def\be{\begin{equation}}
\def\ee{\end{equation}}

\usepackage{amsmath,amssymb}
\usepackage{slashed}
\usepackage{xcolor} 
\usepackage{graphicx}
\usepackage{url}

\newcommand{\beq}{\begin{equation}}
\newcommand{\eeq}{\end{equation}}
\newcommand{\bea}{\begin{eqnarray}}
\newcommand{\eea}{\end{eqnarray}}
\newcommand{\nn}{\nonumber \\ }

\newcommand{\Gev}{{\rm GeV}}
\newcommand{\kev}{{\rm keV}}

\newcommand{\Mpl}{M_{\rm Pl}}
\newcommand{\TRH}{T_{\rm RH}}
\newcommand{\TCMB}{T_{\rm CMB}}
\newcommand{\xiend}{\xi_{\rm end}}
\newcommand{\Hend}{H_{\rm end}}
\newcommand{\HRH}{H_{\rm RH}}
\newcommand{\aRH}{a_{\rm RH}}
\newcommand{\aend}{a_{\rm end}}
\newcommand{\amax}{a_{\rm max}}

\newcommand{\kend}{k_{\rm end}}
\newcommand{\kmax}{k_{\rm max}}
\newcommand{\xend}{x_{\rm end}}
\newcommand{\kast}{k_\ast}
\newcommand{\rhoDend}{\rho^{\rm{end}}_D}
\newcommand{\rhoIend}{\rho^{\rm{end}}_I}

\newcommand{\fref}[1]{Fig.~\ref{fig:#1}} 
\newcommand{\eref}[1]{Eq.\eqref{eq:#1}} 

\newcommand{\sref}[1]{Sec.~\ref{sec:#1}}

\graphicspath{{./Figures/}}
\input{definitions.sty}

\begin{document}

\vspace*{-2cm}
\begin{flushright}
UG-FT 333-21, CAFPE 203-21
\vspace*{2mm}
\end{flushright}

\title{\Large{Dark photon dark matter
from a rolling inflaton }}

\author{\bf {Mar Bastero-Gil}}
\email{mbg@ugr.es}
\affiliation{\normalsize\it
Departamento de F\'{i}sica Te\'{o}rica y del Cosmos and CAFPE,~Universidad de Granada, Campus de Fuentenueva, E-18071 Granada, Spain}

\author{\bf{Jose Santiago}}
\email{jsantiago@ugr.es}
\affiliation{\normalsize\it
Departamento de F\'{i}sica Te\'{o}rica y del Cosmos and CAFPE,~Universidad de Granada, Campus de Fuentenueva, E-18071 Granada, Spain}

\author{\bf{Lorenzo Ubaldi}}
\email{ubaldi.physics@gmail.com}
\affiliation{\normalsize\it
SISSA and INFN Sezione di Trieste,Via Bonomea 265, 34136 Trieste, Italy}
\affiliation{\normalsize\it
Institute for Fundamental Physics of the Universe,
 Via Beirut 2, 34014 Trieste, Italy}

\author{\bf{Roberto Vega-Morales}}
\email{rvegamorales@ugr.es}
\affiliation{\normalsize\it
Departamento de F\'{i}sica Te\'{o}rica y del Cosmos and CAFPE,~Universidad de Granada, Campus de Fuentenueva, E-18071 Granada, Spain}

{
\begin{abstract}\vspace{0.3cm}
We study in detail a recently proposed mechanism for producing non-thermal dark photon dark matter at the end of inflation in the mass range $\mu\,{\rm eV} \lesssim m \lesssim 10\,{\rm TeV}$.~A tachyonic instability induced by a rolling inflaton leads to the \emph{coherent} production of dark (abelian) gauge bosons with a peak in the power spectrum corresponding to the Hubble scale at the end of inflation.~As the Universe expands after inflation the dark photons redshift and, at some point in their cosmic evolution, they obtain a mass.~We focus in particular on the case where the dark photons are \emph{relativistic} at the time their mass is generated and examine the associated cosmic evolution to compute the relic abundance today.~We also examine the late time power spectrum demonstrating explicitly that it preserves the peak generated at the end of inflation.~We show that the peak corresponds to small physical scales today, $\ell_{\rm today} \sim {\rm cm} -  100\,{\rm km}$, with large density fluctuations at $\ell_{\rm today}$ leading to a clumpy nature for the dark photon dark matter.~We also discuss potential phenomenology and future directions, briefly commenting on the non-relativistic case.
\end{abstract}
}

\maketitle
\tableofcontents

\section{{\bf Introduction}} \label{sec:intro} 

The nature and production mechanism of dark matter still remains a mystery.~While most of the experimental effort in the past was aimed at detecting weakly interacting massive particles (WIMPs), the lack of observation has necessitated new theoretical ideas and proposed search strategies to cover as many alternatives as possible~\cite{Bertone:2018xtm}.~In this context, it is important to explore new production mechanisms for dark matter and assess their impact on search strategies.~As the WIMP paradigm comes into increasing tension, candidates for non-thermal dark matter have gained renewed interest~\cite{Bertone:2018xtm}.~Recently, one class of non-thermal dark matter candidates receiving considerable attention is that of vector (or ``dark photon"\,\footnote{We use `dark photon' and `dark vector' interchangeably throughout to refer to a neutral spin-one massive vector boson associated with a broken dark $U(1)_D$ gauge symmetry.}) dark matter for which several production mechanisms have been proposed.~These include production mechanisms associated with inflation~\cite{Nelson:2011sf,Arias:2012az,Graham:2015rva,Bastero-Gil:2018uel,Ema:2019yrd,Nakayama:2020rka,Nakai:2020cfw,Ahmed:2020fhc,Kolb:2020fwh,Salehian:2020asa,Firouzjahi:2020whk} as well as oscillating scalars after inflation~\cite{Agrawal:2018vin,Co:2018lka,Dror:2018pdh}.~The common feature among them is a coupling between a dark abelian gauge boson and a separate sector which induces a time dependence into the dispersion relation of the dark vector field.~This separate sector can come in the form of a scalar like the inflaton~\cite{Bertone:2018xtm,Salehian:2020asa} or an axion~\cite{Agrawal:2018vin,Co:2018lka} or simply gravity~\cite{Graham:2015rva,Ahmed:2020fhc,Kolb:2020fwh}.~Over large regions of parameter space this leads to exponential dark vector production and can reproduce the observed dark matter relic abundance over many orders of magnitude of dark matter mass. 

Motivated by, but not limited to, scenarios of axion inflation, in~\cite{Bastero-Gil:2018uel} it was shown that a dark abelian gauge field coupled to an inflaton via a $\phi F\tilde{F}$ coupling can be produced by a tachyonic instability and generate the observed dark matter relic abundance in the mass range $\mu\,{\rm eV} \lesssim m \lesssim 10\,{\rm TeV}$.~More specifically, the time dependance induced by the rolling inflaton leads to a tachyonic enhancement and exponential production of \emph{one transverse polarization} of the dark photon.~As the Universe expands after inflation the dark gauge bosons redshift and, at some point in their cosmic evolution, they obtain a mass and become non-relativistc.~As in~\cite{Graham:2015rva} where the longitudinal mode is produced by inflationary fluctuations, there is a peaked structure in the power spectrum.~However, in this case the peak is not due to redshifting, but instead to the time dependence of the inflaton as it rolls down its potential.~As we examine in detail below, the dark photon production is exponentially sensitive to the inflaton velocity.~This leads to the maximum production just at the end of inflation as the inflaton exits slow-roll.~This in turn gives rise to a peak in the dark photon power spectrum at scales corresponding to the Hubble scale at the end of inflation. 

The goal of this work is to examine this mechanism in detail and to compute the late time energy density spectrum.~We focus in particular on the case where the dark vector is \emph{relativistic} at the time its mass is generated and examine the associated cosmic evolution.~We first review the production mechanism and track the total energy density to estimate the parameter space which can reproduce the observed dark matter relic abundance.~We then examine the energy density spectrum at the end of inflation as well as its evolution to late times once the dark vector has become non-relativistic.~We obtain the late time power spectrum demonstrating explicitly that the peak generated at the end of inflation is preserved after redshifting.~We  then show that the peak corresponds to small physical scales today, $\ell_{\rm today} \sim {\rm cm} -  100\,{\rm km}$, with potentially large density fluctuations at $\ell_{\rm today}$ leading to a clumpy nature for the dark photon matter.~We also discuss potential phenomenology and future directions, briefly commenting on the non-relativistic case.

\section{{\bf Vector dark matter production from end of inflation}}

Here we expand on the discussion presented in~\cite{Bastero-Gil:2018uel} to show that a dark vector field coupled to a slow-rolling inflaton via a $\phi F\tilde{F}$ coupling can be produced by a tachyonic instability and generate the observed dark matter relic abundance.~To do this we first derive the equations of motion and obtain (approximate) analytic solutions for the tachyonic modes.~We then compute the total energy density at the end of inflation and track its cosmic evolution. 

Since we work in the weak field regime where backreaction effects can be neglected, the evolution of the dark vector modes is linear.~Thus, while the vector field begins as quantum fluctuations during inflation that turn classical only later on, the mode functions of the creation and destruction operators can be obtained using the classical equations of motion.~The evolution of the energy density and power spectrum can then be directly extracted from the mode functions obeying classical evolution.
\subsection{Tachyonic production in an expanding universe }
\label{sec:vdmprod}

Our starting point is the action for an inflaton with potential $V(\phi)$ coupled to a spin-1 vector boson which is neutral under the Standard Model gauge group,
\bea
S  &=& - \int d^4x \, 
\sqrt{-g} \Big[\frac{1}{2} \partial_\mu \phi \partial^\mu \phi + V(\phi) 
+~\frac{1}{4} F_{\mu\nu} F^{\mu\nu} + \frac{1}{2} m^2 A_\mu A^\mu + \frac{\alpha}{4 f} \phi F_{\mu\nu} \tilde F^{\mu\nu} \Big] \, ,
\label{eq:Lag}
\eea
where $\phi$ is the inflaton field that drives inflation, $A_\mu$ is the dark vector field, $F_{\mu\nu}= \partial_\mu A_\nu - \partial_\nu A_\mu$ is the field strength, and $\tilde F^{\mu\nu} = \epsilon^{\mu\nu\alpha\beta} F_{\alpha\beta}/2$ with $\epsilon^{\mu\nu\alpha\beta}$ the completely antisymmetric tensor.~We use the Friedmann-Robertson-Walker metric with $ds^2 = - dt^2 + a^2(t) d\vec{x}^2$ and the convention $\epsilon^{0123} = 1/\sqrt{-g}$.~The vector boson mass $m$ can be zero or non-zero during inflation and can be of a Stueckelberg or `Higgs-ed' type with an associated symmetry breaking phase transition.~As long as it is smaller than the Hubble scale at the end of inflation the vector mass has negligible effects on the tachyonic production mechanism and only becomes relevant when considering cosmological evolution and the final relic abundance.~There could in principle be a kinetic mixing between the visible and dark photons, but it does not spoil the production mechanism as long as it is small enough to prevent thermalization of the dark vector with the visible sector and decay to Standard Model particles for masses above the electron threshold as well as satisfy other experimental constraints~\cite{Gherghetta:2019coi}.~We also do not specify the inflaton potential $V(\phi)$ as its precise form is not crucial for the production mechanism.~However, the form of the inflationary potential can affect the shape of the dark photon energy density spectrum as we examine in more detail in~\sref{energy}.

The only crucial ingredients needed for the production mechanism are:~$(i)$ the scale of the inflaton potential which sets the Hubble scale during inflation,
\bea
H = \frac{\sqrt{V(\phi)}}{\sqrt{3} \Mpl} \, ,
\eea
and $(ii)$ the coupling of the inflaton to $F\tilde F$ responsible for exponential production of only one polarization of the transverse modes.~Such a coupling is generically present in models of natural inflation, where $\phi$ is a pseudoscalar (odd under parity) axion-like field subject to a shift symmetry.~For this reason, this class of models provides a well motivated theoretical framework for the mechanism presented here.~However, since the dynamics of the mechanism do not depend on whether $\phi$ is an axion or not, $\phi$ can be a generic scalar (or function of $\phi$~\cite{Barnaby:2011qe}) as long as it is rolling towards its minimum and couples to $F \tilde F$
\footnote{These ingredients are also present in some models aimed at the relaxation of the electroweak scale \cite{Tangarife:2017rgl} or of the cosmological constant \cite{Graham:2019bfu}.}.~Note that we do not need to impose that the Lagrangian respects parity so $\phi$ could also be parity even and, in particular, is not necessarily a pseudo Goldstone boson.

We quantize the vector field by expanding in the helicity basis in terms of creation and annihilation operators and their mode functions as follows,
\bea
\hat{\vec A}(\vec x,t) &=& 
\sum_{\lambda = \pm,L} \int \frac{d^3k}{(2\pi)^3} 
e^{i \vec k \cdot \vec x} \ \vec \epsilon_\lambda(\vec k) 
\times [A_\lambda(k,t) a_\lambda(\vec k) + A_\lambda(k,t)^\ast a_\lambda^\dagger(-\vec k)] ,
\eea
where we include transverse and longitudinal polarisation in the sum and the creation and annihilation operators satisfy the commutation relation,
\bea
\Big[a_\lambda(\vec k),\,a_\lambda^\dagger(\vec k^\prime) \Big] = (2\pi)^3 
\delta_{\lambda\lambda^\prime}\,\delta^3(\vec k - \vec k^\prime) .
\eea
The mode functions obey the equations of motion derived in the Appendix starting from the action in~\eref{Lag} and which in Fourier space read~\cite{Anber:2009ua,Barnaby:2010vf},
\bea
\ddot \phi &+& 3 H \dot\phi + V'  = 
\frac{\alpha}{4f}  F \tilde F   , \label{eq:EOMphi} \\
\ddot A_\pm &+& H \dot A_\pm + \left( \frac{k^2}{a^2} \pm \frac{k}{a} \frac{\alpha\dot\phi}{f} + m^2  \right) A_\pm  = 0 \, , \label{eq:AEOMT} \\
\ddot A_L &+& \frac{3 k^2 + a^2 m^2}{k^2 + a^2 m^2} H \dot A_L + \left( \frac{k^2}{a^2} + m^2 \right) A_L = 0 \, ,
\label{eq:AEOML}
\eea
where we have also included the inflaton equation of motion.~The overdots denote derivatives with respect to physical time $t$ and $k \equiv |\vec k|$ is the magnitude of the comoving momentum.~We consider only the spatially homogeneous zero momentum mode ($k=0$) of the inflaton.~We have separated the three degrees of freedom of the vector into transverse and longitudinal components, $\vec A_T$ and $A_L$ respectively, where $\vec k \cdot \vec A = k A_L$ and $\vec k \cdot \vec A_T = 0$, and we have written the transverse component in terms of the two helicities, $\vec A_T = \vec\epsilon_+ A_+ + \vec\epsilon_- A_-$.~We see explicitly the $\phi F\tilde F$ coupling only enters into the equations of motion for the transverse modes.~The equation of motion for $A_L$ then corresponds to the one derived in~\cite{Graham:2015rva} and thus, as demonstrated in~\cite{Graham:2015rva}, if $m\neq 0$ during inflation the longitudinal mode is produced via inflationary fluctuations and will also contribute to the dark vector energy density.

In what follows we concentrate on the equation of motion of the transverse vector modes in~\eref{AEOMT}.~It is convenient to introduce the dimensionless ``instability parameter'',
\bea \label{eq:xidef}
\xi \equiv \frac{\alpha \dot \phi}{2H f} = \sqrt{\frac{\epsilon}{2}} \frac{\alpha}{f} \Mpl  \, ,
\eea
where $\epsilon \equiv - \dot H/H^2$ and for single field inflation we have,
\bea\label{eq:slowroll}
|\dot\phi| \approx  V'/3H,~~ 
\epsilon =  \frac{\dot \phi^2}{2H^2 M_{Pl}^2} .
\eea
We then rewrite the equation of motion in terms of conformal time $\tau$ defined as $a d\tau = dt$, 
\bea
\label{eq:AEOMtau}
\Big[ \frac{\partial^{2}}{\partial\tau^{2}} + k^{2} 
\pm 2\,k\,\frac{\xi}{\tau} + \frac{\bar{m}^2}{\tau^2}
\Big] A_\pm(k, \tau)
\equiv
\Big[ \frac{\partial^{2}}{\partial\tau^{2}} 
+ \omega^2(k,\tau) \Big] A_\pm(k, \tau)
= 0 \, ,
\eea
where we have defined the dimensionless ratio,
\bea\label{eq:mbar1}
\bar m \equiv \frac{m}{H} .
\eea
Without loss of generality we use the convention $\dot\phi > 0$ which gives $\xi > 0$, implying that \emph{only} the mode $A_+$ experiences a tachyonic instability when,
\bea \label{eq:Omegatach}
\omega^2(k,\tau) = k^2 + 2k \frac{\xi}{\tau} + \frac{\bar{m}^2}{\tau^2} 
= k^2 - 2 k \xi a H + \bar{m}^2 (a H)^2 
< 0 \, , 
\eea
where we have used the fact that during inflation ($\tau < 0$) we have $\tau \simeq -\frac{1}{a H}$.~On the other hand, the opposite polarization $A_-$~does not have tachyonic modes and is therefore neglected.~The condition in~\eref{Omegatach} then leads to the tachyonic conditions on the vector mass (which could be zero during inflation) and physical momentum ($q$),
\bea\label{eq:qtac}
~~~~
q \equiv
\frac{k}{a} < 
\xi H + \xi H \sqrt{1- \bar{m}^2/ \xi^2} ~~\rm{~~~(tachyonic~condition)},
\eea
which requires $\bar{m} < \xi$ to avoid the mass term quenching the tachyonic production~\cite{Meerburg:2012id}.~When $\bar{m} \ll \xi$, we also see the tachyonic condition on the physical momentum becomes $q < 2\xi H$.~As we will see below, we are interested in the (weak coupling) regime where $\xi$ is $\mathcal{O}(1)$ implying that the modes become tachyonic as their physical wavelength is stretched to be of order the horizon, $\lambda \equiv q^{-1} \sim H^{-1}$.~Since $-\omega^2$ is maximal at $q \simeq \xi H$, this implies the vector field power spectrum should also have a peak at scales around the co-moving horizon.~Within the horizon these modes add up coherently and have very large occupation number due to the exponential enhancement from the tachyonic instability.~Thus, these dark vectors are well described by a classical dark `electromagnetic' field and since only one helicity is enhanced exponentially, we are left at the end of inflation with a maximally \emph{helical dark electromagnetic field}.~This is in analogy with magnetogenesis scenarios~\cite{Adshead:2016iae} constructed to solve the puzzle of the origin of primordial magnetic fields. 

Treating $\xi$ as a constant, a good approximation early on during inflation as the inflaton slow-rolls, one can solve~\eref{AEOMtau} analytically~\cite{Meerburg:2012id} in terms of the Whittaker functions.~The overall normalization is determined by the requirement that the gauge field is initially (in the sub-horizon limit $-k\tau \to \infty$) in the Bunch-Davies (BD) vacuum
\footnote{This choice of normalization makes the classical mode solutions consistent with the quantum ones.},
\bea\label{eq:ABD}
\lim_{-k\tau\to \infty} A_\pm (k,\tau) = \frac{e^{-ik\tau}}{\sqrt{2k}} \equiv A_{\rm{BD}} \, .
\eea
Neglecting the dark photon mass (with $\xi$ as constant) the full analytic solution to~\eref{AEOMtau} can be found in terms of Coulomb functions~\cite{Meerburg:2012id,Adshead:2015pva,Domcke:2018eki} which, in the tachyonic regime we are interested in $-k\tau < 2\xi$ ($k < 2 \xi a H$), are very well approximated by~\cite{Barnaby:2011vw},
\bea \label{eq:Atachsol}
A_+(k,\tau) &\simeq & e^{\pi \xi}  \sqrt{\frac{-2 \tau}{\pi}} K_1 \left[ 2 \sqrt{-2\xi k \tau} \right] \, ,
\eea
where $K_1$ is a modified Bessel function of the second kind.~In this form we see the exponential dependence on $\xi $ (or $\dot{\phi}$) explicitly.~In the super-horizon (SH) limit ($-k\tau\to 0$)~\eref{Atachsol} gives,
\bea \label{eq:ATP}
\lim_{-k\tau\to 0} A_+ (k,\tau) = \frac{e^{\pi \xi}}{2 \sqrt{2 \pi k \xi}} \equiv A_{\rm{SH}} \, .
\eea
Useful analytic solutions can be obtained via the WKB approximation~\cite{Anber:2009ua, Barnaby:2011vw, Tangarife:2017rgl},
\bea \label{eq:AWKB}
A_+(k,\tau)_{\rm WKB} & \simeq & \frac{1}{\sqrt{2k}} \left( \frac{-k\tau}{2\xi} \right)^{1/4} e^{\pi \xi - 2 \sqrt{-2\xi k\tau}} ~~~~
(\frac{1}{8\xi}  <   -k\tau  < 2\xi) \, ,
\eea
where the regime of validity is dictated by the adiabatic condition $|\Omega' / \Omega^2| \ll1$ with $\Omega' \equiv d\Omega/d\tau$.~For modes with $k$ in the range $aH/(8\xi) < k < 2\xi aH$, $A_+(k,\tau)_{\rm WKB}$ approximates very well the solution obtained in~\eref{Atachsol} and gives us intuition into the behavior of the modes around horizon crossing as they become exponentially enhanced.~This also allows us to use analytic solutions to obtain an estimate of the total vector dark matter relic abundance and the viable regions of parameter space for the mechanism.~Eventually, we will need to compute the power spectrum at the end of inflation to use as input when tracking the cosmological evolution of the energy density spectrum.~However, as we emphasize in ~\sref{energy}, a more precise calculation of the dark matter energy density and shape of the power spectrum requires accounting for the time dependence of $\xi$ which necessitates solving the system of equations in~\eref{EOMphi} and~\eref{AEOMT} numerically.

\subsection{Energy density at the end of inflation}
\label{sec:prod}

To eventually obtain the final relic abundance for the vector dark matter we need to track the evolution of the energy density starting from the time of its production at the end of inflation.~Thus we need to first compute the total vector energy density at the end of inflation.~As we review in the Appendix, starting from the action in~\eref{Lag} we can obtain the total energy density for the transverse component of the dark vector field ($\rho_D$) in terms of the tachyonic mode amplitude and its (conformal) time derivative,
\bea 
\rho_D
&=&
\frac{1}{4\pi^2 a^4} \int_0^\infty dk \, k^2 
\Big( | \partial_\tau A_{+}(k,\tau) |^2 + 
\left( k^2 + a^2 m^2 \right) |A_{+}(k,\tau)|^2 \Big) \nn
&=& \frac{1}{2a^4} \int d\ln k 
\Big( \mathcal{P}_{\partial_\tau A_+}(k,\tau) + \left( k^2 + a^2 m^2 \right) \mathcal{P}_{A_+}(k,\tau) \Big) \nn
&=& \langle (\partial_\tau A_+)^2 \rangle + \langle A_+^2 \rangle \,  
= \frac{1}{2}\langle \vec{E}^2 + \vec{B}^2 \rangle.
\label{eq:rhoDint}
\eea
where $\rho_D \equiv \langle \rho_D \rangle$ represents the spatial average as defined in the Appendix (see~\eref{rhoTt}) and we identify the `magnetic' ($B$) and `electric' ($E$) components respectively.~We have used $dt = a d\tau$ as well as defined the field and (time) derivative power spectra,
\bea\label{eq:PSdef}
\mathcal{P}_X (k,\tau) &=& \frac{k^3}{2\pi^2} |X (k,\tau) |^2~~~
X = A_{+} \ {\rm or} \ \ \partial_\tau A_{+} \, ,
\eea 
which allows us to define the electric and magnetic energy density spectra respectively,
\bea \label{eq:rhoEandB}
\frac{d\rho_E}{d\,{\rm ln}\,k} &=& 
\frac{1}{2a^4} 
 \mathcal{P}_{\partial_\tau A_+}(k,\tau) ,~~
\frac{d\rho_B}{d\,{\rm ln}\,k} = 
\frac{1}{2a^4} 
\left( k^2 + a^2 m^2 \right)
 \mathcal{P}_{A_+}(k,\tau) \,.
\eea
Note as part of the magnetic component we have included the mass term which of course is not present in the case of the visible electromagnetic field.~Since only one transverse mode is exponentially enhanced (which we take as $A_+$) by the tachyonic instability, we can safely neglect the contribution from $A_-$ to~\eref{rhoDint}.~For the case where the dark vector already has a mass during inflation, we also compute the energy density contained in the longitudinal mode and obtain the same result as in~\cite{Graham:2015rva}.

Taking $\bar{m} \ll 1$ during inflation and neglecting the time dependence of $\xi$, we can use the WKB solution for $A_+$ in~\eref{AWKB} to compute analytically the dark electric field contribution given by $\mathcal{P}_{\partial_\tau A_+}(k,\tau)$ as well as the dark magnetic field contribution $k^2 \mathcal{P}_{A_+}(k,\tau)$, where the former always dominates over the latter during inflation.~Integrating over momenta we can then estimate the energy density contained in the dark vector during inflation,
\bea\label{eq:rhoDinf}
\rho_D 
\approx 10^{-4} \frac{e^{2\pi \xi}}{\xi^3} \,  H^4 .
\eea
In reality of course, both the Hubble parameter and $\xi$ depend on time.~The parameter $\xi$, which controls the dark photon production and grows with $\dot\phi$ (or the slow-roll parameter $\epsilon$), is largest at the end of inflation.~Thus, the largest contribution to the dark electromagnetic energy density comes from the end of inflation, as confirmed in our numerical analysis discussed below.~From \eref{rhoDinf} we can estimate the energy density at the end of inflation as,
\bea \label{eq:rhoD}
\rho_D(a_{\rm{end}})  \equiv \rhoDend 
\approx 10^{-4} \frac{e^{2\pi \xiend} }{\xiend^3} \Hend^4 \, 
= 10^{-4} \frac{e^{2\pi \xiend}}{\xiend^3}  \, \epsilon_H^4 H^4 ,
\eea
with $\xiend$ the value of $\xi$ at the end of inflation and given by (using~\eref{xidef}),
\bea\label{eq:xiend}
\xiend = \frac{\alpha}{\sqrt{2}}\frac{\Mpl}{f} ,
\eea
where we have assumed $\epsilon = 1$ at the end of inflation\footnote{Note in hybrid inflation models~\cite{Copeland:1994vg,Dvali:1994ms}, one can have values different from $\epsilon \approx 1$ at the end of inflation.}.~In the last equality in~\eref{rhoD} we account for a decreasing Hubble parameter during inflation and parametrize it as,
\bea \label{eq:Hend}
\Hend = \epsilon_H H \, ,
\eea
with $\epsilon_H$ a dimensionless parameter that can be calculated in a particular model of inflation.~Typically the slow-roll parameter $\epsilon = - \dot H/H^2$ is $\mathcal{O}(10^{-2} - 10^{-3})$ during inflation and $\approx 1$ at the end of inflation.~This translates into values of $\epsilon_H$ that are model dependent, but for most models of inflation we expect $\epsilon_H$ to be in the range $10^{-3} < \epsilon_H < 10^{-1}$.~In axion inflation models involving tachyonic production of vector fields, which have roughly 60 e-folds of inflation, $\xi$ during the first few e-folds is constrained by CMB measurements~\cite{Barnaby:2010vf,Barnaby:2011vw} to be less than $\xi_{\rm CMB} \lesssim 2.5$, but $\xi_{\rm end}$ is allowed to be significantly larger.~However, if $\xi$ is too large this will induce back reaction effects which must be accounted for and which could potentially destroy the production mechanism.~In the slow-roll regime where analytic approximations of the amplitude can be used this imposes the constraint $\xi < 4.7$~\cite{Barnaby:2011qe,Peloso:2016gqs}, but near the end of inflation where the analytic approximations over estimate the amplitude and numerical solutions must be used (see~\fref{energyspeccomp}) we find $\xi < \mathcal{O}(10)$.~Thus, in order to neglect these back reaction effects we limit ourselves to $\xi_{\rm end} < \mathcal{O}(10)$ in the following.~On the other hand, to obtain sufficient dark vector production requires $\xi_{\rm end} \gtrsim 1$ leading us to consider the range $1 \lesssim \xi_{\rm end} \lesssim 10$ when examining the viable vector dark matter parameter space below.
 
During inflation, when $\phi$ dominates the energy density, we have for the inflaton,
\bea \label{eq:rhoI}
\rho_I = V(\phi) = 3 H^2 \Mpl^2 \, ,
\eea
while at the end of inflation, with the inflaton still dominating, we have $H = \Hend$ and,
\bea \label{eq:rhoIend}
\rhoIend = 3 \Hend^2 \Mpl^2 =
3 \, \epsilon_H^2 H^2 \Mpl^2 \, .
\eea
A fraction of $\rhoIend$ is transferred into the dark vector while another fraction goes into visible radiation to reheat the Universe whose energy density we can write as,
\bea \label{eq:rhoRH}
\rho_R (T_{\rm RH})  
= \frac{\pi^2}{30}g_*(\TRH) \TRH^4 \, 
\equiv 
\epsilon_R^4 \, \rho_I 
= 3 \HRH^2 \Mpl^2 .
\eea
Combining with~\eref{rhoI}, this allows us to define the reheating temperature,
\bea \label{eq:TRH}
\TRH = \epsilon_R\, \left( \frac{90}{\pi^2 g_*(\TRH)} \right)^{1/4} \sqrt{H \Mpl} \, ,
\eea
as well as the Hubble scale at reheating in terms of the Hubble scale during inflation,
\bea \label{eq:HHR}
\HRH = \epsilon_R^2 H \, .
\eea
The dimensionless parameter $\epsilon_R < 1$ parametrizes the fraction of the inflaton energy that goes into visible radiation and in general can take on values spanning many orders of magnitude depending on the particular model of reheating.~We take $g_*(\TRH)$ to denote the number of relativistic degrees of freedom which we fix to $g_*(\TRH) \sim 100$ and restrict ourselves to reheating temperatures above the electroweak scale.~We work in the approximation of instantaneous reheating assuming it takes place as soon as the inflaton exits slow-roll.~Thus, we assume the Universe has a temperature $\TRH$ at $a = \aend$ (see~\fref{confdiag}) implying $\rhoDend = \rho_D(\TRH)$.~Note however, we do not assume $\HRH = \Hend$ which will be important to account for when we discuss constraints on the vector dark matter parameter space below.

\subsection{Evolution of total energy density} \label{sec:EDevolution}

In this section we track the redshift of the dark vector energy density after production at the end of inflation and estimate the present day relic abundance.~Here we only consider the case where the dark vector is relativistic at the time its mass is generated.~We also assume the dark vector mass is already present during inflation and remains non-zero throughout its cosmic evolution.~This scenario can be applied to either a Stueckelberg or dark Higgs mechanism for generating the dark vector mass.~We study the non-relativistic case and other possibilities for cosmic evolution when the dark vector mass is generated via a dark Higgs mechanism in~\cite{Higgsfollowup} where we find a different result from~\eref{DMT1} for the relic density.~For purposes of tracking the evolution and estimating the viable dark matter parameter space, it is enough to track the redshift of the energy density considering only modes around the peak of the power spectrum for which we assume they all redshift together.

As discussed in~\sref{vdmprod}, for a given scale factor, the power spectrum will be peaked around scales the size of the co-moving horizon.~Thus, at the end of inflation the modes which give the largest contribution to $\rhoDend$ have physical momentum,
\bea\label{eq:qRH}
q(\TRH) \equiv \frac{k}{\aend} \simeq \Hend \gg m\, ,
\eea
where again we have assumed instantaneous reheating allowing us to define $\TRH$ and track the redshift using temperature instead of scale factor.~At reheating the dark vectors are relativistic with the physical momentum then redshifting as,
\bea\label{eq:qT}
q(T) = q(\TRH) \frac{T}{\TRH} \, .
\eea
The dark photons become non relativistic at a temperature $T = \bar T$ defined by the condition,
\bea\label{eq:qeqm}
q(\bar T) = m \, , 
\eea
which combining with~\eref{Hend},~\eref{TRH} and~\eref{qRH} allows us to solve for $\bar{T}$ as,
\bea
\bar T 
= \frac{m}{\Hend} \, \TRH
= m  \left( \frac{90}{\pi^2 g_*(\TRH)} \right)^{1/4} \frac{\epsilon_R}{\epsilon_H}   \left( \frac{\Mpl}{H} \right)^{1/2} \, .
\label{eq:Tbar}
\eea
Above $\bar T$ (or before scale factor $a = \bar{a}$) the vector energy density redshifts like radiation, 
\bea \label{eq:rhoDrad}
\rho_D(T) = \rho_D(\TRH) \left( \frac{T}{\TRH} \right)^4 \, ,
\eea
while below $\bar T$ (after $a = \bar{a}$) it redshifts like matter giving,
\bea \label{eq:rhoDmat}
\rho_D(T) = \rho_D(T_0) \left( \frac{T}{T_0}  \right)^3 \, ,
\eea
where $T_0 \approx 10^{-13}$ GeV is today's CMB temperature.~The cosmic evolution described above and in the previous section can be summarized with~\fref{peakevo} which shows the evolution of the various energy densities defined in~\eref{rhoD},~\eref{rhoI}, and~\eref{rhoRH}.
\begin{figure}[tbh]
\includegraphics[scale=.25]{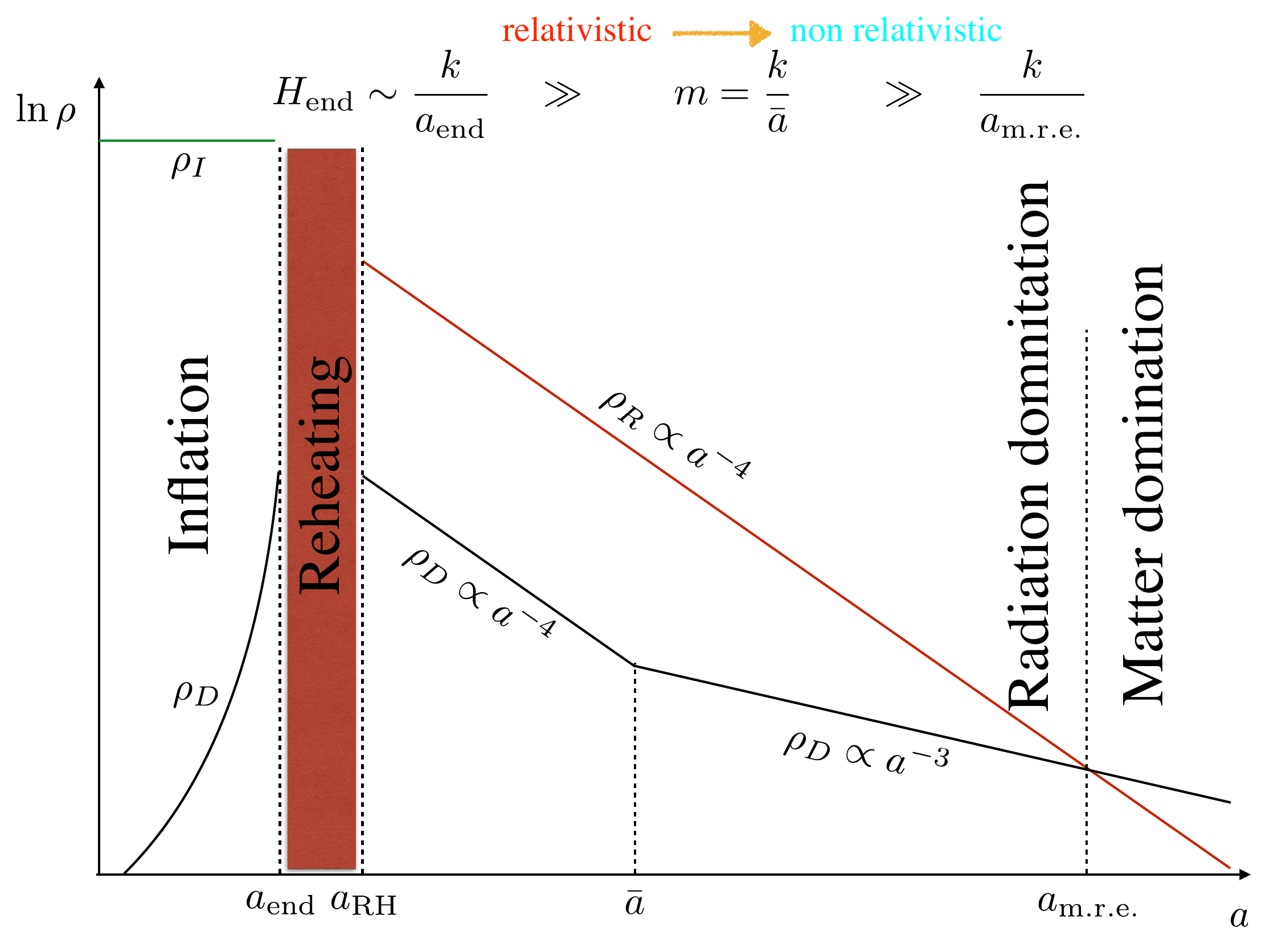}
\caption{Schematic representation of the cosmic evolution of the vector dark matter energy density ($\rho_D$) as well as the energy density in visible radiation ($\rho_R$) and the inflaton ($\rho_I$).}
\label{fig:peakevo}
\end{figure}

Equating~\eref{rhoDrad} and~\eref{rhoDmat} at $T = \bar{T}$ and combining with~\eref{rhoD},\,\eref{TRH},\,and~\eref{Tbar}, we obtain the energy density today in terms of the energy density at the end of inflation,
\bea
\rho_D(T_0) &=&
m \,T_0^3
 \left( \frac{90}{\pi^2 g_*(\TRH)} \right)^{-3/4} 
\left( \frac{\epsilon_H}{ \epsilon_R} \right)^3 
\left( \frac{H}{\Mpl } \right)^{3/2} 
\left(\frac{\rhoDend}{\Hend^4}\right) \, .
\label{eq:DMeq1}
\eea
This expression assumes the modes around the peak all redshift together which is strictly speaking not true as modes of different momenta become non-relativistic at different times, but for present purposes the approximation in~\eref{DMeq1} is sufficient.~The dark vector energy density at the end of inflation $\rhoDend$ (normalized to $\Hend^4$) can be obtained numerically, but as found in~\eref{rhoD} a useful analytic result can be obtained with the WKB approximation. 

Using the WKB approximation and taking the observed energy density of cold dark matter today~\cite{Ade:2015xua} to be $\rho_{\rm CDM} = 9.6 \times 10^{-48} \ {\rm GeV}^4$, the contribution from the transverse dark vector mode produced via tachyonic instability can then be written as,
\bea
\frac{\rho_D(T_0)}{\rho_{\rm CDM}} =
\frac{\Omega_T}{\Omega_{\rm CDM}} \simeq
2 \times 10^{-4} \cdot
\left( \frac{m}{\rm MeV}  \right)
\left( \frac{H}{10^{14} \ {\rm GeV}} \right)^{3/2}  
\left( \frac{e^{2\pi \xiend}}{\xiend^3} \right) \,
\left( \frac{\epsilon_H}{\epsilon_R} \right)^3 .
\label{eq:DMT1}
\eea
We see in this case the final relic abundance depends on five parameters.~The first two are the dark vector mass $m$ and the Hubble scale during inflation $H$, as in the case of the longitudinal mode~\cite{Graham:2015rva}.~The third parameter $\xiend$ parametrizes the strength of the inflaton - dark vector coupling (see~\eref{xidef}) and depends weakly on the precise shape of the inflaton potential.~The final two parameters parametrize our ignorance of the inflaton sector, $\epsilon_H$ which parametrizes how much $H$ has decreased by the end of inflation (see~\eref{Hend}) and $\epsilon_R$ which parametrizes the fraction of energy transferred from the inflaton to the reheating sector (see~\eref{rhoRH}).~Both can in principle be calculated in a particular model of inflation.

If the dark vector has a mass already during inflation, whether of Stueckelberg or Higgs-ed type (with the $U(1)_D$ \emph{not} restored during reheating), which persists throughout the entirety of its cosmic evolution until today, there is also a contribution to the relic density from the longitudinal mode produced via inflationary quantum fluctuations~\cite{Graham:2015rva},
\bea \label{eq:DML}
\frac{\Omega_L}{\Omega_{\rm CDM}} =
\left( \frac{m}{ 6 \times 10^{-15} \ {\rm GeV}} \right)^{1/2}  \left( \frac{H}{10^{14} \ {\rm GeV}} \right)^2 \, .
\eea
We see this only depends on the vector mass and Hubble scale during inflation.~In this case both the transverse and longitudinal modes could in principle contribute appreciably to the dark matter relic abundance.~Thus, for these cases of dark vector mass generation, we will also include the longitudinal mode when exploring the viable parameter space below.

\subsection{Vector dark matter parameter space} \label{sec:paramspace} 

Having followed the cosmic evolution of the the dark vector energy density and found an estimate for the relic abundance today, we can go on to estimate the regions of parameter space for a viable dark matter candidate.~To find these regions, we must ensure that not only the observed cold dark matter relic abundance is reproduced, but also that a number of constraints on the parameters in~\eref{DMT1} are satisfied for consistency of the mechanism.~As discussed in the previous section, we will focus only on scenarios where the dark vector is relativistic at the time its mass is generated and assume in the Higgs-ed case that once broken, the dark $U(1)_D$ is not restored by reheating or any other phase transition~\cite{Higgsfollowup}.

\begin{center}
{\bf \emph{Constraints on model parameters}}
\end{center}

\begin{enumerate}

\item For the constraints on the dark vector mass the upper bound results from requirng efficient tachyonic production at the end of inflation and constrains only the case of $m \neq 0$ during inflation.~At the same time $m$ is bounded from below by the condition that the dark vector becomes non-relativistic, thus behaving like cold dark matter, before matter-radiation equality.~We write this condition as $\bar T > \TCMB$ (see~\eref{Tbar}), with $\TCMB \simeq 10^{-9}$ GeV.~Thus $m$ is constrained to lie in the window,
\bea \label{eq:mconstraint}
\left(\frac{\pi g_*(\TRH)}{90}\right)^{1/4} \frac{\epsilon_H}{\epsilon_R} \sqrt{\frac{H}{\Mpl}} \, \TCMB < 
m &<&
\epsilon_H H\, .
\eea  
This window allows for dark vector masses spanning many orders of magnitude.
\item The energy density in the inflaton at the end of inflation must be larger than in the radiation which reheats the Universe, $\rhoIend > \rho_R(\TRH)$ implying $\Hend > \HRH$ and,
\bea \label{largepsH}
\frac{\epsilon_H}{\epsilon_R^2} > 1 \, ,
\eea
to ensure the Universe does not reheat to energies greater than $\rhoIend$.

\item At $a = \aRH = \aend$, the energy density of radiation in the reheating sector must be greater than the energy density of the dark vector, $\rho_R (\TRH) >  \rhoDend$ which leads to,
\bea \label{eq:epsilonconstraint}
 \frac{\epsilon_H}{\epsilon_R}  \ll  10 \  \xiend^{3/4} e^{-\pi\xiend/2} \left( \frac{\Mpl}{H} \right)^{1/2} \, .
\eea
Otherwise, the Universe would become (dark) matter dominated at a temperature above $\TCMB$, thus violating matter-radiation equality at $\TCMB$. 
\item If the inflaton - dark vector coupling is too large, the dark vector can thermalize with the inflaton which must also couple to Standard Model particles to reheat the Universe.~This would lead to thermalization of the dark vector with the visible sector and spoil our dark matter production mechanism.~Ensuring the inflaton and dark vector do not thermalize~\cite{Ferreira:2017lnd, Ferreira:2017wlx} puts an upper bound on $\xi$,
\bea
\xi < 0.44 \ln \frac{f}{\alpha H} + 3.4 \, .
\eea
Using~\eref{xidef} and taking the slow-roll parameter to be $\epsilon = 1$ at $\aend$ then gives,
\bea \label{eq:thermal}
\xiend < 0.44 \ln \left( \frac{1}{\sqrt{2} \xiend} \frac{\Mpl}{H}  \right) + 3.4 \, .
\eea

\item We also assume that back-reaction effects on the inflaton dynamics are negligible which leads to two conditions.~The first is $3 H \dot\phi \simeq V' \gg \alpha / f \langle \vec E \cdot \vec B \rangle$ meaning that the $\langle F\tilde{F}\rangle = \langle \vec E \cdot \vec B \rangle$ term is negligible in the inflaton equation of motion in~\eref{EOMphi}.~The second condition is that $3 H^2 \Mpl^2 \gg \langle \vec E^2 \rangle / 2$, ensuring that the inflaton dominates the energy density during inflation rather than the dark vector.~Both conditions are satisfied as long as $\xi$ is not too large~\cite{Barnaby:2011qe,Peloso:2016gqs}.~Requiring that they hold all the way to the end of inflation results in the constraint on $\xiend$,
\bea \label{eq:backreaction}
\frac{\epsilon_H H}{\Mpl} \ll 10^2\, \xiend^{3/2} \, e^{-\pi\xiend} \, .
\eea
\item When the inflaton exits the slow-roll regime, it starts oscillating about the minimum of its potential and reheats the Universe.~If the coupling $\alpha / f$ is moderately large, roughly $\alpha / f > 35 \Mpl^{-1}$, the production of dark vectors during these oscillations can be important.~This phenomenon is referred to as gauge-preheating and has been studied in~\cite{Adshead:2015pva}.~However, in order to satisfy the constraints listed above, here we consider the range $1 \lesssim \xi_{\rm{end}} \lesssim 10$ which results in the window for $\alpha/f$,
\bea
\frac{\sqrt{2}}{M_{\rm{Pl}}} \lesssim \frac{\alpha}{f} \lesssim 
\frac{10 \sqrt{2}}{M_{\rm{Pl}}}  \, ,
\eea
where we have used~\eref{xiend}.~In this range of $\alpha / f$, preheating into dark vectors is not efficient~\cite{Adshead:2015pva} implying that during the inflaton oscillations only a negligible fraction of its energy density is transferred to the dark vector.~We then assume that reheating proceeds via the perturbative decay of the inflaton into visible radiation. 

\end{enumerate}

Finally, we also implicitly assume there are no other light scalar or fermion fields in the dark sector which couple to the dark vector.~If such light fields were present, they would be produced by the strong dark electromagnetic field via the Schwinger effect~\cite{Tangarife:2017vnd, Tangarife:2017rgl} and would in principle contribute to the dark matter abundance today.~Here we assume they are heavy enough to avoid this interesting possibility which we leave to forthcoming work~\cite{DSfollowup}.

\begin{center}
{\bf \emph{Final viable parameter space}}
\end{center}

In~\fref{relic} we show the relic abundance given in~\eref{DMT1} as a function of $m$ and $H$, imposing the constraints listed above, for different values of the parameters $\xiend$, $\epsilon_R$, $\epsilon_H$.~In practice, requiring the dark vector is not over abundant along with the first two constraints automatically ensures the remaining constraints are satisfied, so only these are shown.~Along the contours labeled ``Transverse'' for different values of $\xiend$ which separate the purple shaded regions we obtain the observed relic abundance with the transverse mode making up the entirety of the dark matter.~In the colored regions to the right of these lines the dark matter is overabundant.~We see that the transverse mode of the dark vector can make a viable dark matter candidate over a wide range of parameter space:
$\mu {\rm eV} \lesssim m \lesssim {\rm TeV}$,
$100 \ {\rm GeV} \lesssim H \lesssim 10^{14} \ {\rm GeV}$ for $\xiend \sim \mathcal{O}(1-10)$ (see~\eref{xiend}).
\begin{figure}[tbh]
\includegraphics[scale=.62]{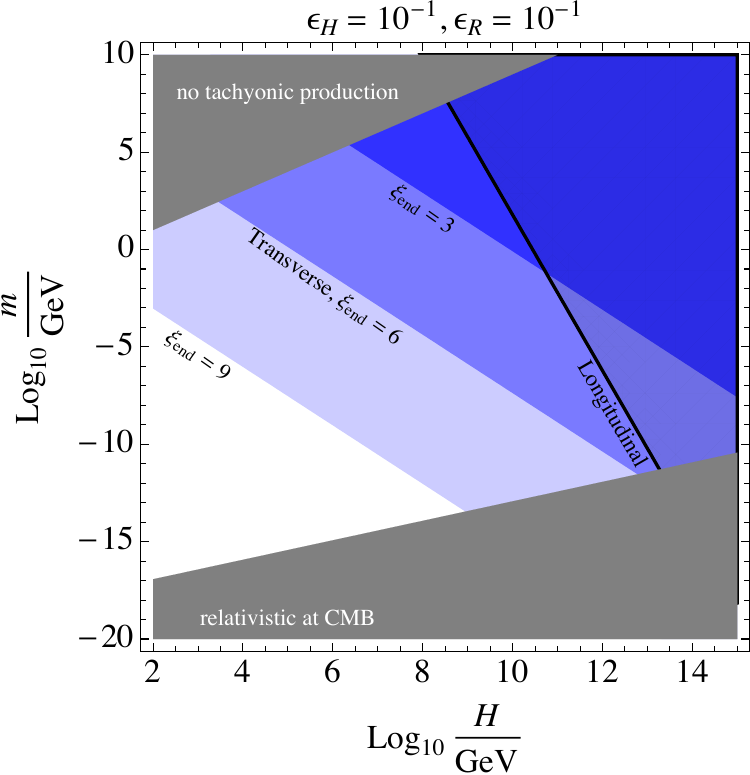}
\includegraphics[scale=.62]{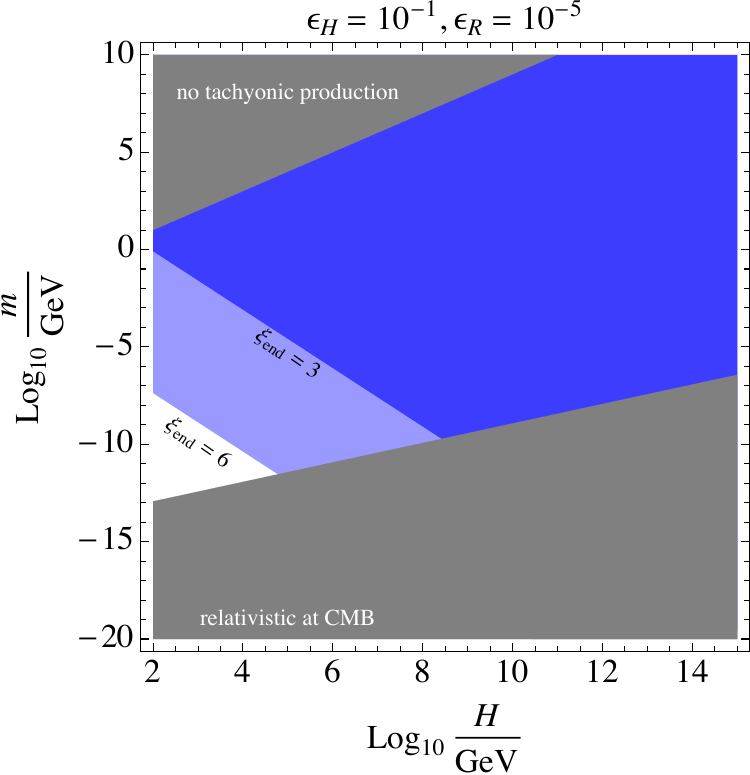}
\caption{Parameter space in the dark photon mass versus Hubble scale $m - H$ plane for values of the parameters $\xiend$, $\epsilon_R$, and $\epsilon_H$ as indicated above each plot and described in the text.~Along the lines labeled ``Transverse'' for different values of $\xiend$, we obtain the observed relic abundance for the transverse mode while in the colored regions to the right of these lines the dark matter is overabundant.~The region in the grey band at large masses is excluded by requiring efficient tachyonic production during inflation while the region in gray at low masses is excluded by requiring the dark photons are non-relativistic by the time of CMB formation (see~\eref{mconstraint}).~For the case when the dark vector has a mass during inflation, we also show along the black line labeled ``Longitudinal'' the contour where the longitudinal mode makes up all of the observed dark matter.}
\label{fig:relic}
\end{figure}

For comparison, we also plot the relic abundance of the longitudinal mode in~\eref{DML} in the case the dark vector has a mass during inflation and is thus also produced via inflationary fluctuations~\cite{Graham:2015rva}.~In the regions where the line labelled `Transverse' is to the left of the ones labelled `Longitudinal', the transverse mode gives the dominant contribution to the relic density.~We see large regions of parameter space where this is the case.~On the left in~\fref{relic} we also see a region of parameter space where the longitudinal and transverse modes give comparable contributions to the dark matter relic abundance which raises interesting possibilities to be discussed more below.~Since for the parameters shown on the right in~\fref{relic} the transverse mode always dominates we do not show the contour for the longitudinal mode.~Note however the relic abundance for the longitudinal mode only depends on $m$ and $H$ so it has the same contour regardless of the other parameters.

Finally, for illustration purposes we consider a specific benchmark point,
\bea\label{eq:bench}
\xiend=6\, , \ \ \epsilon_R = 10^{-1}\, , \ \ \epsilon_H=10^{-1}\, , \\
H = 10^9 \ \Gev\, , \ \ m = 1.3 \ \kev\, .\nonumber
\eea
This leads to a reheating temperature $\TRH = 2.7 \times 10^{12}$ GeV and an initial radiation energy density $\rho_R(\TRH) = 1.7 \times 10^{51} \ \Gev^4$, several orders of magnitude larger than the initial energy density in the dark electromagnetic field $\rho_D(\TRH) = 10^{42} \ \Gev^4$.~The dark photons become non-relativistic at $\bar T = 36$~MeV, then redshift like matter for some time before matching the energy density of radiation at $\TCMB$.~Note that the momentum of the dark photon has a long time to redshift from $\bar T$ to $\TCMB$ so it is very `cold' by the time of matter-radiation equality.~Note also that at the time of Big Bang Nucleosynthesis (BBN), $T_{\rm BBN} \sim 1$ MeV, the dark photon is already non-relativistic and still constitutes a small fraction of the total energy density.~Therefore, bounds on extra relativistic species ($N_{\rm eff}$) are easily avoided.

\section{{\bf Energy Density Spectrum and Clumpy Dark Matter}} \label{sec:energy} 
 
Here we examine in detail the cosmological evolution of the dark vector energy density starting from the end of inflation.~In particular, we examine the energy density spectrum and show explicitly that the peak in the spectrum at the end of inflation survives cosmic evolution  until late times.~We then examine the spectrum of \emph{fluctuations} in the energy density around the time of matter radiation equality.~We confirm that power at large scales remains highly suppressed allowing for a vector produced in this manner to evade constraints on isocurvature from measurements of the CMB~\cite{Akrami:2018odb}, thus making it a viable dark matter candidate.~The power spectrum of density fluctuations also serves as the starting point for studying implications on structure formation~\cite{Kolb:1994fi,Graham:2015rva,Alonso-Alvarez:2018tus,Berges:2019dgr} which we leave for future work.

\subsection{Energy density spectrum at the end of inflation} \label{sec:PSend} 

To gain further intuition for the tachyonic production mechanism and evolution of the modes during inflation as well as facilitate the discussion below, we can consider the analytic solutions in~\eref{ABD}-\eref{AWKB} together with the conformal diagram in~\fref{confdiag}.~Here we show co-moving length scales versus scale factor (or conformal time) with the co-moving horizon indicated by the contour (solid black) at $k^{-1} = (a H)^{-1}$.~The last mode to exit the horizon during inflation is indicated by the black line labeled $k_{\rm{end}}^{-1} = (a_{\rm{end}} \Hend)^{-1}$.~The Compton wavelength contour at $\lambda = q^{-1}  = (k/a)^{-1} = m^{-1}$ applies in the case that the dark vector already has a mass during inflation and defines the time when it becomes non-relativistic.~We also show the last mode to cross this contour (and become non-relativistic) during inflation labeled $k_m$ as well as the maximum momentum tachyonic mode $\kmax \simeq 2\xi \kmax$.~Utilizing~\fref{confdiag} and~\eref{qtac} we find the ratios of these scales,
\begin{figure}[tbh]
\includegraphics[scale=.5]{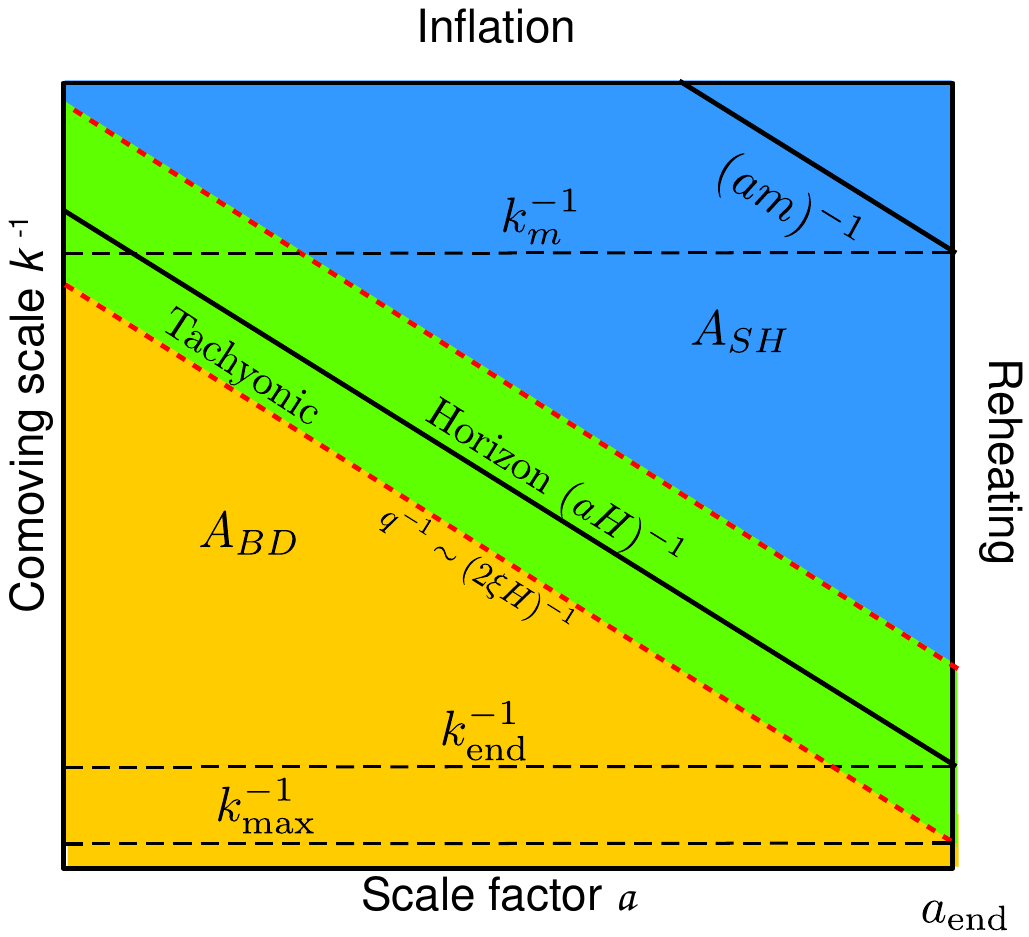}
\caption{Conformal diagram during inflation zoomed in around the horizon region (green) showing co-moving scales versus scale factor (see text for more information).}
\label{fig:confdiag}
\end{figure}
%
\bea\label{eq:scaleratios1}
\frac{k_m}{k_{\rm{end}}} &=& \frac{m}{\Hend},~~
\frac{k_{\rm{max}}}{k_{\rm{end}}} = 2\xiend + \mathcal{O}(\frac{m^2}{\Hend^2}),
\eea 
where $\Hend$ indicates the Hubble scale at the end of inflation.~In the super horizon limit ($-k\tau\to 0 $) shown in blue the amplitudes approach the asymptotic solution in~\eref{ATP}.~In the orange region deep inside the horizon ($-k\tau \to \infty$) the modes are in the Bunch-Davies vacuum given by the solution in~\eref{ABD}.~As the modes enter the horizon region shown in green and~\eref{qtac} is satisfied, the tachyonic instability leads to exponential growth of the amplitude for one of the transverse modes (which we take to be $A_+$).~As we'll see in~\sref{PSend}, for a fixed point in time (or scale factor), the power in the dark electromagnetic field is dominated by modes contained within this region leading to a peak in the power spectrum at $k \sim a H$ when the modes have a wavelength of order the horizon.~This leads to a power spectrum at the end of inflation that is peaked at co-moving scales around $k_{\rm{end}}^{-1} $.

The result for the total dark vector energy density at the end of inflation obtained in~\eref{rhoD} relied on the WKB approximation for the amplitude in~\eref{AWKB} which assumed the time dependence of $\xi$ could be neglected.~This was sufficient for obtaining an estimate of the relic density and viable dark matter parameter space in~\fref{relic}.~However, the instability parameter $\xi \propto \dot\phi \propto \sqrt\epsilon$ is not only largest towards the end of inflation, but also experiences the largest \emph{growth} just as inflation is ending and the slow-roll parameter approaches $\epsilon \approx 1$ (in single field inflation scenarios).~Since the energy density depends exponentially on $\xi$ (see~\eref{rhoD}), to obtainin an accurate density spectrum it is crucial to account for this time dependence.~However, to account for the time dependence of $\xi$ and the Hubble parameter as well as the breakdown of the slow-roll approximation, it is necessary to numerically solve the equations of motion in~\eref{EOMphi} and~\eref{AEOMT}.~This requires a 
robust integration procedure which we describe in the Appendix.

Having performed this numerical integration, we show in~\fref{energyspeccomp} a comparison between the energy density spectrum obtained numerically (solid) versus analytically in~\eref{Atachsol} (dashed) for both the electric (blue) and magnetic (red) components (see~\eref{rhoEandB}).~We consider a $\phi^4$ type inflaton potential and $m/\Hend \ll 1$.~On the left we show the spectra at early times during inflation as the CMB modes leave the horizon where we require $\xi$ to satisfy $\xi_{\rm{CMB}} < 2.5$~\cite{Barnaby:2010vf,Barnaby:2011vw}.~On the right we show the same spectra, but now just at the end of inflation for $\xiend = 9$.~As we can see, at early times during inflation when $\xiend \approx 1$ the analytic and numerical solutions are in good agreement.~However, at the end of inflation we see the shape, location, and exact height of the peak depends not just on $\xiend$ (which is the same in both cases), but on how $\xi$ changes with time.~We also see that the time dependence in $\xi$ leads to stronger suppression of power at large scales as well as a lower peak than the analytic case.~This allows us to consider larger values of $\xiend$ before backreaction effects become important as compared to the analytic approximation for the amplitude which holds only during the slow-roll regime where the time dependence of $\xi$ can be neglected~\cite{Barnaby:2011qe,Peloso:2016gqs}.~Note this way of suppressing power at large scales is distinct from other dark matter production mechanisms connected to inflation which also lead to a peaked power spectrum~\cite{Graham:2015rva,Alonso-Alvarez:2018tus,Berges:2019dgr}. 
\begin{figure}[tbh]
\begin{center}
\includegraphics[scale=.36]{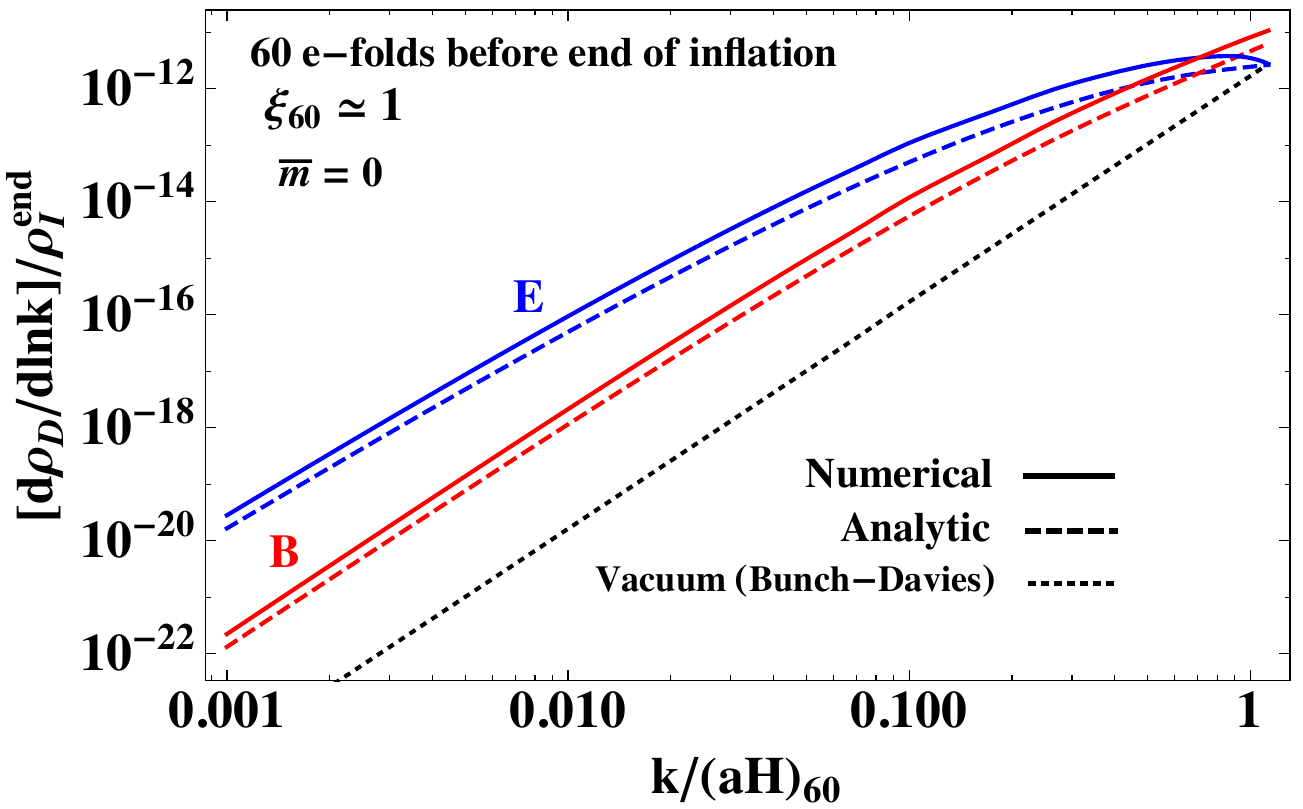}
\includegraphics[scale=.36]{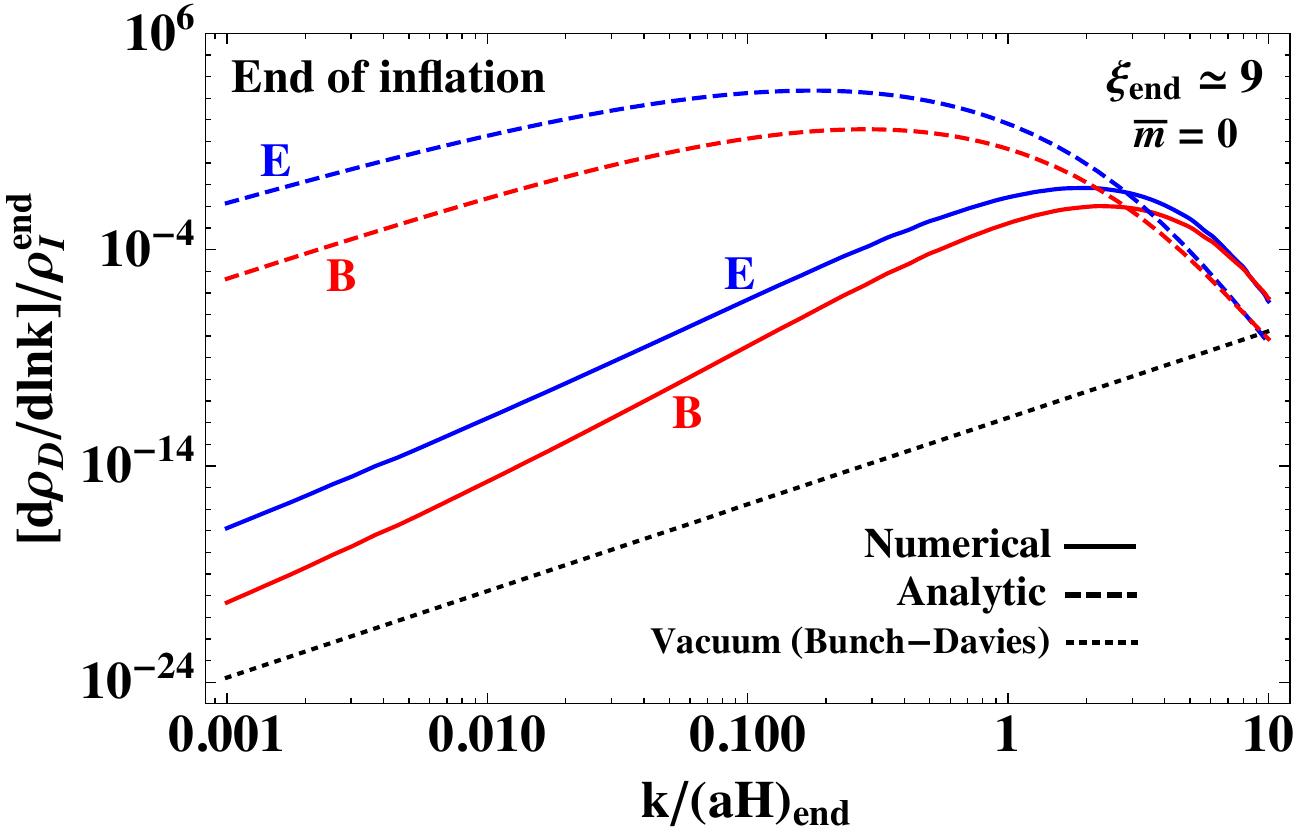}
\end{center}
\caption{Energy density spectrum for the electric (E) and magnetic (B) components in~\eref{rhoEandB} obtained numerically (solid) versus analytically (dashed) at both early times during inflation (left) as the CMB modes leave the horizon and just at the end of inflation (right) for $\xiend = 9$ and $m = 0$.~The spectrum for the Bunch-Davies vacuum modes is also shown (dotted).}
\label{fig:energyspeccomp}
\end{figure}

With the numerical solutions in hand, we obtain the energy density spectrum at the end of inflation for various inflationary scenarios.~We first examine the effects of the dark vector having a mass already during inflation\footnote{In this case there may be a constraint from the Swampland conjecture which requires $m > 60$\,eV~\cite{Bastero-Gil:2018uel,Reece:2018zvv}.} which can arise through either a Stueckelberg or Higgs mechanism.~In~\fref{energyspec} we show the energy density spectrum for the electric (solid) and magnetic (dashed) component at the end of inflation as a function of $k/k_{\rm{end}}$ for $m/\Hend = 0~(\rm{black}),~3.15\cdot10^{-5}~(\rm{red}),~3.15\cdot10^{-1}~(\rm{green})$ and $\xiend = 9$.~The spectrum for the Bunch-Davies vacuum (black dotted) is also shown.~We see all the spectra are above the vacuum for modes with $k/a_{end} < 2\xiend \Hend$ showing the particle production effect.~Higher momentum modes stay in the Bunch-Davies vacuum and have a spectrum below the vacuum one once a proper subtraction scheme has been implemented (see Appendix).~This signals the absence of the tachyonic instability and particle production effects for these modes.
\begin{figure}[tbh]
\begin{center}
\includegraphics[scale=.32]{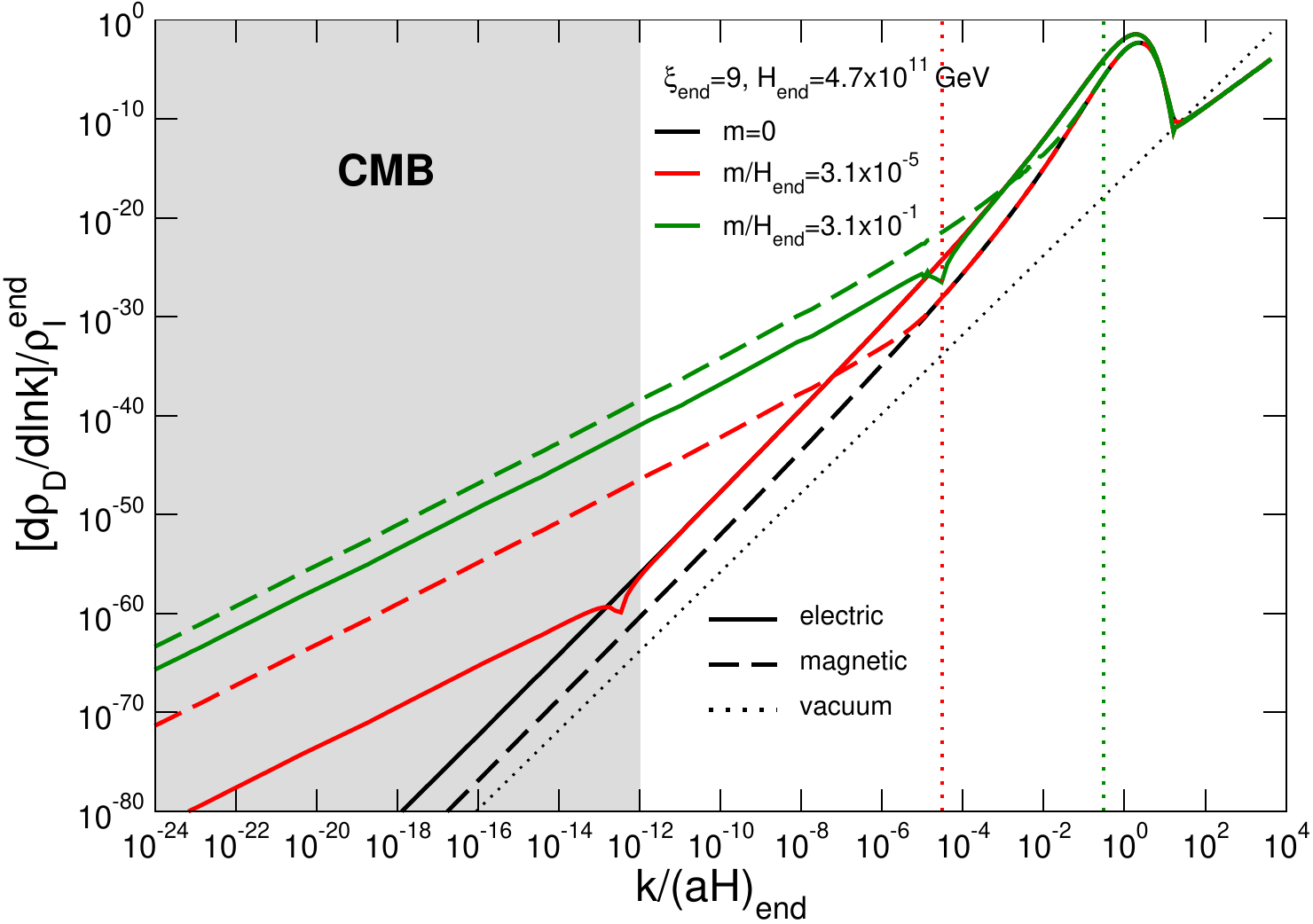}
\includegraphics[scale=.32]{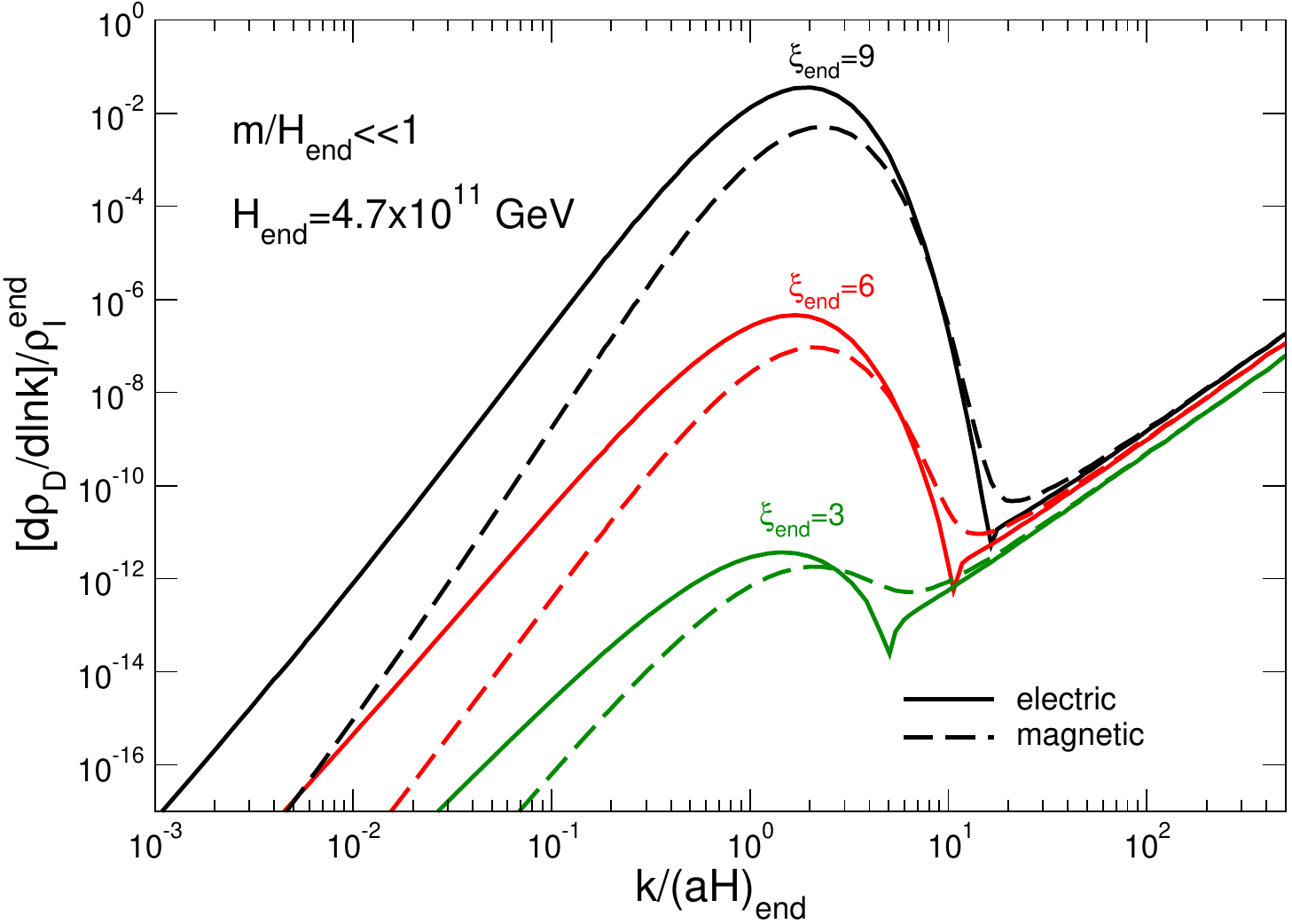}
\end{center}
\caption{{\bf Left:}\,Energy density spectrum for the electric (solid) and magnetic (dashed) component at the end of inflation for $m/\Hend = 0~(\rm{black}),~3.15\cdot10^{-5}~(\rm{red}),~3.15\cdot10^{-1}~(\rm{green})$ and $\xiend = 9$.~The red and green dotted vertical lines indicate $k_m/k_{\rm{end}} = m/\Hend$ (see~\fref{confdiag} and~\eref{scaleratios1}).~The spectrum for the Bunch-Davies vacuum (black dotted) is also shown.~{\bf Right:}\,Same energy density spectra as left, but zoomed in around the peak for $\xiend = 3, 6, 9$ and $m/\Hend \ll 1$.}
\label{fig:energyspec}
\end{figure}

We also see a number of features which arise when the dark vector has a mass already during inflation.~For instance, we see at around $k_m/k_{\rm{end}} = m/\Hend$ (see~\fref{confdiag} and~\eref{scaleratios1}) indicated by the red and green dotted vertical lines, the slope in the spectrum for the magnetic component changes from decreasing like $k^{-4}$  to one decreasing like $k^{-2}$ as we go to larger scales.~This change occurs when $q = m$ and the modes at large co-moving scale (top in~\fref{confdiag}) cross the Compton wavelength contour in~\fref{confdiag} causing them to become non-relativistic and damp more slowly with expansion than the still relativistic modes at smaller scales.~As discussed above, the last mode for which this occurs is at $k_m$ so modes at scales larger than $k_m^{-1}$ will see an enhancement relative to the massless case as seen in~\fref{energyspec}.~We see also that at the end of inflation this leads to domination by the magnetic component at scales $k^{-1} > k_m^{-1}$ where $m >> q$.~At even larger scales $k^{-1} \gg k_m^{-1}$, we see the electric component also changes slope from one decreasing like $k^{-4}$  to one decreasing like $k^{-2}$ after going through a kink when the field time derivative $\partial_\tau A_+$ changes sign.~However, we see it still remains subdominant to the magnetic component at these scales.~This is in contrast to the massless case in which the electric component dominates at all scales\footnote{We discuss these mass effects on the spectrum at large scales in more detail in the Appendix.}.~Finally, we see that around the peak the mass effects are negligible and, in particular, the massive and massless cases have the same spectrum for modes $k > k_m$ which will always contain the majority of the peak.~This is the case unless $\Hend \lesssim m < \xi \Hend$ which we do not consider since the tachyhonic production begins to be suppressed.~However, this super heavy mass case could be an interesting possibility.

We next examine the dependence of the energy density spectrum on $\xiend$.~Since around the peak the electric component always dominates and largely does not depend on the vector mass, we can focus on the region around the peak and consider different values for $\xiend$.~On the right in~\fref{energyspec} we show the same spectra as on the left, but only around the peak for $\xiend = 3, 6, 9$ and $m/\Hend \ll 1$.~Here the exponential sensitivity to $\xiend$ becomes clear as well as the domination of the electric component of the energy density.~Note also that only for the $\xiend = 9$ case does the energy density in the dark vector begin to approach the energy density of the inflaton ($\frac{d\rho_E}{d\,{\rm ln}\,k} \approx \rhoIend$) so we can safely neglect back reaction effects. 

\begin{figure}[tbh]
\begin{center}
\includegraphics[scale=.32]{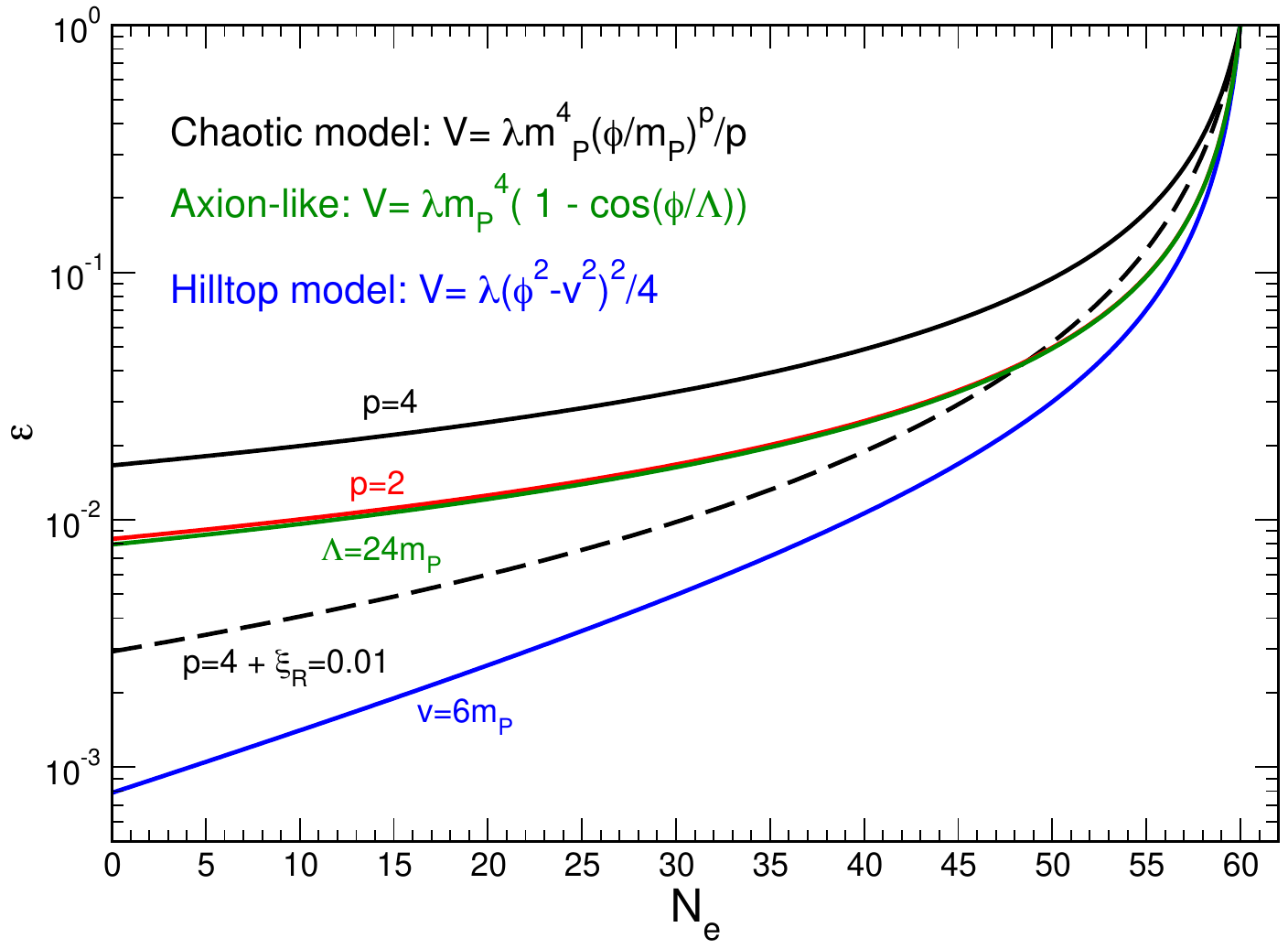}
\includegraphics[scale=.32]{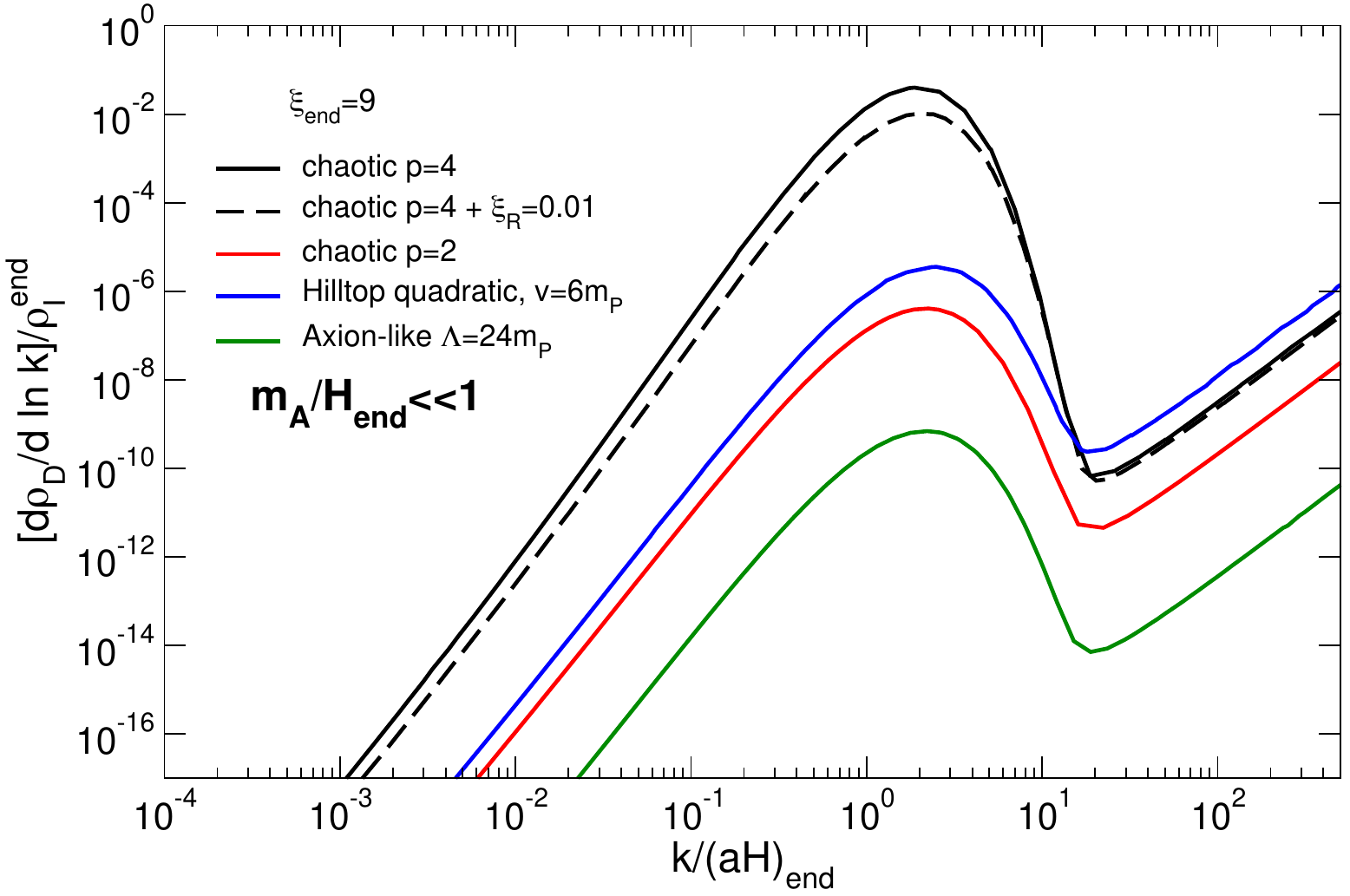}
\end{center}
\caption{{\bf Left:~}\,Slow-roll parameter $\epsilon$ as a function of number of e-folds for various models of inflation.~{\bf Right:}\,Energy density spectrum at the end of inflation for the same models with $\xiend = 9$.}
\label{fig:energyspecsteep}
\end{figure}
 
Finally, we examine how the shape of the inflaton potential affects the dark vector energy density spectrum.~In~\fref{energyspecsteep} we show on the left how the slow-roll parameter $\epsilon$ changes as a function of the number of e-folds for various models of inflation while the energy density spectrum at the end of inflation is shown on the right for the same models.~Since the instability parameter $\xi$ is proportional to $\epsilon$ (see~\eref{xidef}) we see that regardless of the behaviour of $\xi$ early on during inflation, the largest growth occurs just at the very end of inflation as $\xi$ approaches $\xiend$.~We also see that different inflationary potentials can lead to different spectra for the same $\xiend$.~This opens the possibility that by precisely measuring the dark vector power spectrum, we can potentially infer properties of the inflaton potential, but we leave this to future work\footnote{Of the potentials chosen in~\fref{energyspecsteep}, only the non-minimal $\phi^4$ potential (dashed line) is strictly speaking still allowed by data~\cite{Okada:2010jf}, but these curves are only for illustrative purposes.}.~These energy density spectra will serve as the input needed to track the cosmic evolution of the dark vector energy density.~Below we study the cosmic evolution to obtain the late time energy density spectra and its fluctuations.

\subsection{Cosmological evolution of modes}
\label{sec:evo}

Here we track the evolution of the tachyonic modes starting from the end of inflation through to matter radiation equality following a similar analysis to that found in~\cite{Graham:2015rva,Alonso-Alvarez:2018tus}, keeping track of both the (dark) magnetic and electric components as the Universe continues to expand.~As we saw above, when  produced at the end of inflation, the electric component dominates over the magnetic.~However, as we will see below, the magnetic component quickly grows relative to the electric after inflation and `catches up' by the time the Hubble scale becomes comparable to the dark matter mass.~After this point they redshift together like radiation and then eventually like matter by the time of matter radiation equality.  

The cosmological evolution of the field as a function of scale factor or time (either physical or conformal) can be parametrized in the same way as,
\bea\label{eq:PSlate}
\mathcal{P}_X (k,x) &=& \mathcal{P}_X (k,x_{\rm{end}}) \frac{|X(k,x)|^2}{|X(k,x_{\rm{end}})|^2} ,
~~
X = A_+,\,\partial_\tau A_+  ,~~x = a, t, \tau,
\eea 
where the power spectrum is defined in~\eref{PSdef} (with $\tau \to x$) and $x_{\rm{end}}$ indicates the end of inflation.~The input amplitude (plus derivative) and power spectrum are taken at $\xend$ and obtained by numerically solving the equations of motion in~\eref{AEOMtau} as discussed in~\sref{PSend} and the Appendix.~The late time modes can similarly be obtain numerically, but it becomes computationally intensive to evolve them to late times.~An approximate analytic solution for the late time amplitude and its derivative can be found in different limits of the equations of motion in~\eref{AEOMtau}.~These different limits can be depicted geometrically via the conformal diagram shown in~\fref{confdiaglate} where each limit corresponds to one of the five colored regions as indicated.~As we discuss in more detail below, taking the amplitude (plus derivative) and power spectrum at the end of inflation as input, we can obtain approximations to $X(k,x)$ in the various limits and then `glue' them together to construct the full late time amplitude as well as the mean energy density and fluctuation spectrum~\cite{Graham:2015rva,Alonso-Alvarez:2018tus}.
\begin{figure}[tbh]
\begin{center}
\includegraphics[scale=.4]{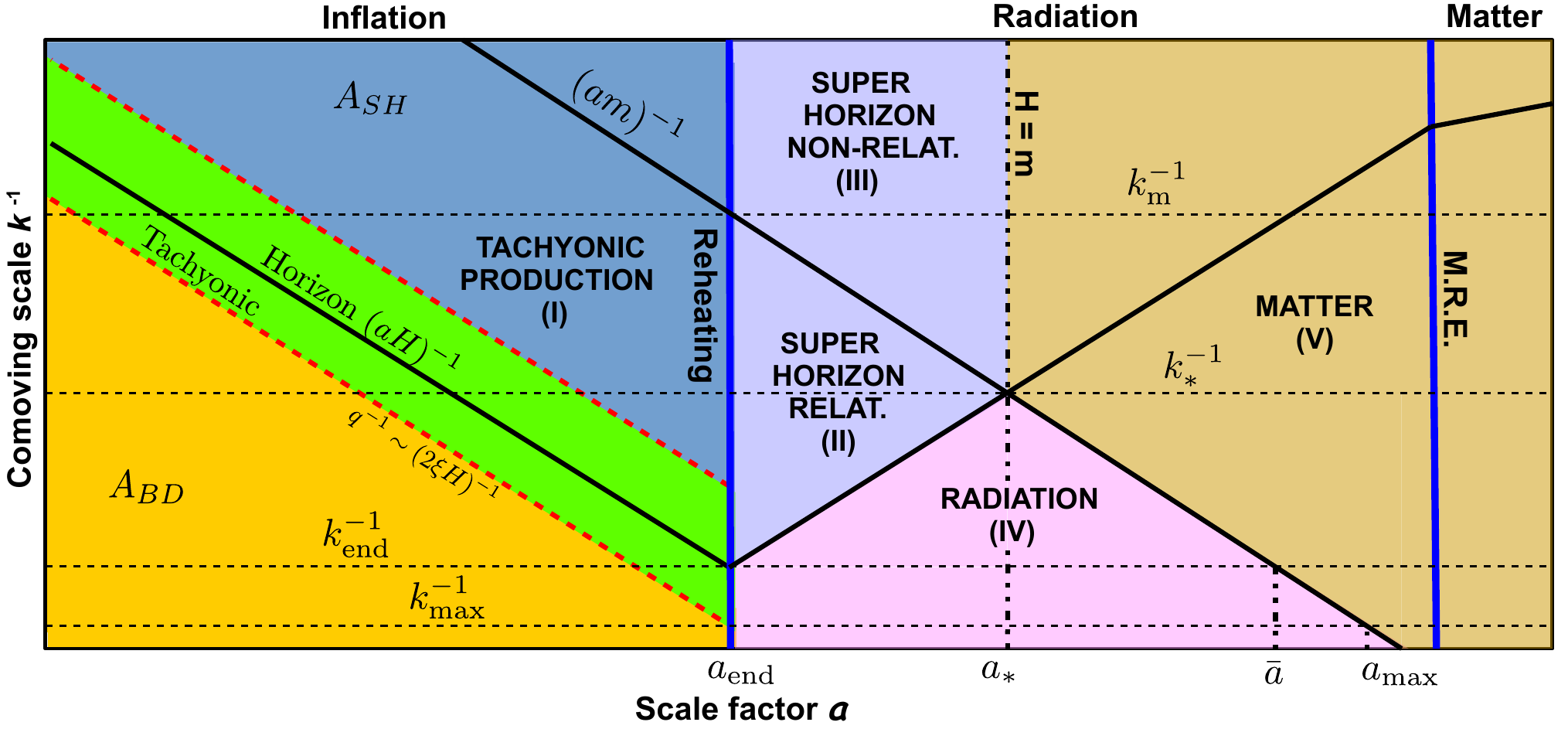}
\end{center}
\caption{Conformal diagram showing co-moving scales versus scale factor from early during inflation until matter radiation equality (see text for more information).}
\label{fig:confdiaglate}
\end{figure}

Using~\fref{confdiaglate} we derive a second set of ratios of scales (in addition to~\eref{scaleratios1}),
\bea\label{eq:scaleratios2}
\frac{k_\ast}{k_{\rm{end}}} = 
\frac{a_{\rm{end}}}{a_\ast} &=& \sqrt{\bar{m}},~~
\frac{a_{\rm{end}}}{\bar{a}} = \bar{m},~~
\frac{a_{\rm{end}}}{a_{\rm{max}}} = \frac{\bar{m}}{2\xiend},
\eea 
where $a_\ast$ is defined as the scale factor when $H = m$ and the $k_\ast$ mode\footnote{As we discuss below, the peak in the power spectrum of the longitudinal mode occurs at $k_\ast.$} becomes non-relativistic (crosses the Compton wavelength contour in~\fref{confdiaglate}) while again $\aend$ is the scale factor at the end of inflation.~The scale factors $\bar{a}$ and $\amax$ (or e-folds) indicate respectively the time when the modes $\kend$ and $\kmax$ become non-relativistic.~Since we are taking the end of inflation as the initial condition for cosmological evolution, unless otherwise stated we define $\bar{m}$ as the ratio of the dark vector mass to Hubble scale at the end of inflation,
\bea\label{eq:mbdef}
\bar m = \frac{m}{\Hend}.
\eea
For the following discussion it will also be useful to rewrite the electric and magnetic energy density spectra respectively as a function of scale factor $a$,
\bea \label{eq:drhoEB}
\frac{d\rho_E}{d\,{\rm ln}\,k}
&=& 
\frac{1}{2} 
\left(\frac{a_{\rm{end}}}{a}\right)^4 
H_{\rm{end}}^2  
\mathcal{P}_{\partial_a A_+}(k,a),~~ 
\frac{d\rho_B}{d\,{\rm ln}\,k} 
=
\frac{1}{2a^2} \left( \frac{k^2}{a^2} + m^2 \right) \mathcal{P}_{A_+}(k,a) 
 \, ,
\eea
where we have used $da = a^2 H d\tau$ and $H = (a_{\rm{end}}/a)^2 H_{\rm{end}}$ for a radiation dominated era.

\begin{itemize}

\item {\bf Region I - End of Inflation: $ H = \Hend$}

During inflation (region I in~\fref{confdiaglate}) quantum fluctuations of the dark vector boson are amplified by the expansion of the Universe leading to the tachyonic instability and exponential production of one polarization.~As discussed, this production is maximal just at the end of inflation leading to a power spectrum peaked at $\kend$ that serves as input for cosmological evolution after inflation (see~\eref{PSlate}).~Explicitly, at scale factor $a = \aend$, we take the tachyonically enhanced amplitude and its derivative to be,
\bea\label{eq:AIdef}
A_{\rm{I}} &=& A_+(k, a_{\rm{end}}),\\
\partial_a A_{\rm{I}} &=& \partial_a A_+(k, a)|_{a = a_{\rm{end}}} \nonumber,
\eea
where $A_+$ and $\partial_a A_+$ are obtained numerically as discussed in~\sref{PSend}.~These will serve as the initial conditions for the cosmological evolution of the tachyonic modes.


\item {\bf Region II - Super-horizon radiation era relativistic: $ H \gg q \gg m $}

Just after inflation we are in the super horizon regime $H \gg q,m$ for modes with $a m < k < \kend$ (region II in~\fref{confdiaglate}).~At this point, the inflaton energy has been converted into reheating and thus, $\xi = 0$ in the equation of motion in~\eref{AEOMT}.~For these relativistic Hubble damped modes with $q \gg m$ this gives an equation of motion,
\bea
\label{eq:EOMII}
(\partial_t^2 + H \partial_t )\,A_+ \simeq 0,
~~~\Longleftrightarrow~~~
(\partial_a a H \partial_a + H \partial_a )\,A_+ \simeq 0 ,
\eea
where we have used $\dot{a} = a H$ and $da = a^2 H d\tau$.~Defining $A_{\rm{II}}$ as the approximate solution for the amplitude in this regime, during the radiation dominated era when the Hubble parameter scales as $H\propto a^{-2}$ we have, 
\bea\label{eq:AII}
A_+ \simeq
A_{\rm{II}} &=& c^{\rm{II}}_1 + c^{\rm{II}}_2 a ,\\
\partial_a A_{\rm{II}} &=& c^{\rm{II}}_2 ,\nonumber
\eea
where $c^{\rm{II}}_{1,2}$ are constants in time, but functions of scale $k^{-1}$, and we see a term in the amplitude that grows linearly with scale factor.~Using~\eref{drhoEB}, we see the electric and magnetic energy density scale as,
\bea
\frac{d\rho_E}{d\,{\rm ln}\,k} \, \propto a^{-4} ,~~
\frac{d\rho_B}{d\,{\rm ln}\,k} \, 
\propto \frac{q^2}{a^2} (c^{\rm{II}}_1 + c^{\rm{II}}_2 a )^2
\propto a^{-(2-4)}  ,
\eea
where again $q = k/a$ is the physical momentum.~However, imposing continuity of the amplitude and derivative at $a_{\rm{end}}$ we also have,
\bea
c^{\rm{II}}_1 &=& A_{\rm{I}} - \aend (\partial_a A_{\rm{I}}) \nn
c^{\rm{II}}_2 &=& \partial_a A_{\rm{I}},
\eea
where we have used~\eref{AIdef}.~So we see the size of the term that grows linearly with $a$ depends on the the size of the input amplitude derivative at the end of inflation.~Since this depends on $k$ and has a peak at $\sim k_{\rm{end}}$, modes at small scales around $\kend$ have a large $c^{\rm{II}}_2$.~However, they spend less time growing linearly before re-entering the horizon while the converse is true for modes at larger scales (see~\fref{confdiaglate}).~The net effect is a brief period of damping like $a^{-2}$ which is much slower than the $a^{-4}$ damping of the electric component.~As we show in~\fref{lateenergyspec} and discuss more below, this brief period of enhancement during super-horizon evolution allows the magnetic component of the energy density to quickly `catch up' to the electric component after which they redshift together.~~Note that in the case of the longitudinal mode the factor of 3 in front of the $\partial_t H$ Hubble damping term leads instead to a solution of the form $A_L \simeq c_1 + c_2 a^{-1}$, which quickly leads to a constant as the Universe expands.


\item {\bf Region III - Super-horizon radiation era non-relativistic: $ H \gg m \gg q $}
 
If the dark vector has a mass already during inflation or one is generated while some modes are still super-horizon during radiation era, we have the possibility of non-relativistic Hubble damped evolution (region III in~\fref{confdiaglate}).~Here we have the same equation of motion as in II leading to the same solutions, including the coefficients ($c^{\rm{III}}_i = c^{\rm{II}}_i$) which at these scales $k \ll \kend$ are very small.~Note, this solution has the same form as the longitudinal case which has the same equation of motion in this region with differing coefficients due to the different input spectrum at the end of inflation.~Unlike in region II, the mass term in~\eref{drhoEB} now dominates over the momentum term for the magnetic component of the dark vector energy density spectrum.~The electric and magnetic energy densities then damp as, 

\bea
\frac{d\rho_E}{d\,{\rm ln}\,k} \, \propto a^{-4} ,~~
\frac{d\rho_B}{d\,{\rm ln}\,k} \, 
\propto \frac{m^2}{a^2} (c^{\rm{II}}_1 + c^{\rm{II}}_2 a )^2
\propto a^{-(0-2)} .
\eea
Since we are far from the peak, these contributions to the energy density are negligible unless $m \sim \Hend$ which we do not consider as we assume $m \ll \Hend$.


\item {\bf Region IV - Sub-horizon radiation era relativistic: $ q \gg m, H $}
 
This is the region in~\fref{confdiaglate} containing the modes around the peak in the dark vector energy density spectrum.~In this regime we have for the equation of motion,
\bea
\label{eq:EOMIV}
\Big(\partial_t^2 + H \partial_t + \frac{k^2}{a^2} \Big)\,A_+ \simeq 0 
~~~\Longleftrightarrow~~~
\Big(\partial_\tau^2 + k^2 \Big)\,A_+ \simeq 0 ,
\eea
where we have used $\partial_\tau a^{-1} = -H$.~Note this differs from the equation of motion for the longitudinal component which in conformal time has a $2a H\partial_\tau$ `damping' term~\cite{Graham:2015rva}.~This is due to a factor of 3 in front of the $H \partial_t$ Hubble damping term in the physical time equation of motion.~Thus~\eref{EOMIV} has solution for the transverse mode,
\bea
A_+ \simeq A_{IV} &=& (c^{\rm{IV}}_1 e^{ik\tau} + c^{\rm{IV}}_2 e^{-ik\tau} ) \nn
\partial_a A_{IV} &=&  \frac{ik}{a_{\rm{end}}^2 H_{\rm{end}}} (c^{\rm{IV}}_1  e^{ik\tau} - c^{\rm{IV}}_2 e^{-ik\tau} ),
\label{eq:AIV}
\eea
which we see has no overall damping.~The longitudinal mode amplitude on the other hand has an overall $a^{-1}$ suppression and the derivative damps like $a^{-2}$.~Note that the factor of $\aend^2 \Hend$ comes from the Jacobian in going from $\partial_\tau$ to $\partial_a$.~For the transverse mode the electric and magnetic energy density then scale as,
\bea
\frac{d\rho_E}{d\,{\rm ln}\,k} \, \propto a^{-4} ,~~
\frac{d\rho_B}{d\,{\rm ln}\,k} 
\propto a^{-2} q^2 \propto a^{-4}.
\eea
We see that the magnetic and electric components of the energy density redshift the same with scale factor in this regime.~Note that while the amplitude scales differently, the damping of the energy density is the same as for the longitudinal mode~\cite{Graham:2015rva}.


\item {\bf Region V - Non-relativistic massive regime: $ m \gg H, q $}

In the non-relativistic massive regime we have $m \gg H,q$ which gives for the equations of motion (now in terms of physical time,~see~\eref{AEOMT}),
\bea
\label{eq:EOMV}
\Big(\partial_t^2 + H \partial_t + m^2 \Big)\,A_+ \simeq 0 .
\eea
This is the same as for the longitudinal component~\cite{Graham:2015rva} and has solution,
\bea\label{eq:AV}
A_+ \simeq A_{V} &=& 
\frac{1}{\sqrt{a}} (c^{V}_1 e^{imt} + c^{V}_2 e^{-imt} ) \nn 
\partial_a A_{V}  &=& 
\frac{im \sqrt{a} }{a_{\rm{end}}^2 H_{\rm{end}}} 
(c^{V}_1 e^{imt} -c^{V}_2 e^{-imt} ) ,
\eea
where we have used the change of variables from physical time to scale factor,
\bea
\label{eq:ttoa}
t \to \frac{1}{2H_{\rm{end}}} (\frac{a^2}{a_{\rm{end}}^2} - 1) + t_{\rm{end}} .
\eea
We see the amplitude damps like $a^{-1/2}$ while the derivative grows like $a^{1/2}$ leading to a scaling for the electric and magnetic energy densities respectively,
\bea
\frac{d\rho_B}{d\,{\rm ln}\,k} &\propto& 
\frac{ m^2}{a^3} \propto a^{-3},~~
\frac{d\rho_E}{d\,{\rm ln}\,k} \propto 
\frac{\aend^4\Hend^2}{a^3} \, \propto a^{-3}  .
\eea
Thus both the magnetic and electric components of the energy density redshift like matter at late times.~In this regime the longitudinal component has the same equation of motion and therefore same solution as in~\eref{AV}.~So again we have the same damping behaviour with scale factor~\cite{Graham:2015rva}, but with different coefficients in~\eref{AV}. 

\end{itemize}

\subsection{Late time energy density spectrum} \label{sec:lateespec} 

\begin{figure}[tbh]
\begin{center}
\includegraphics[scale=.37]{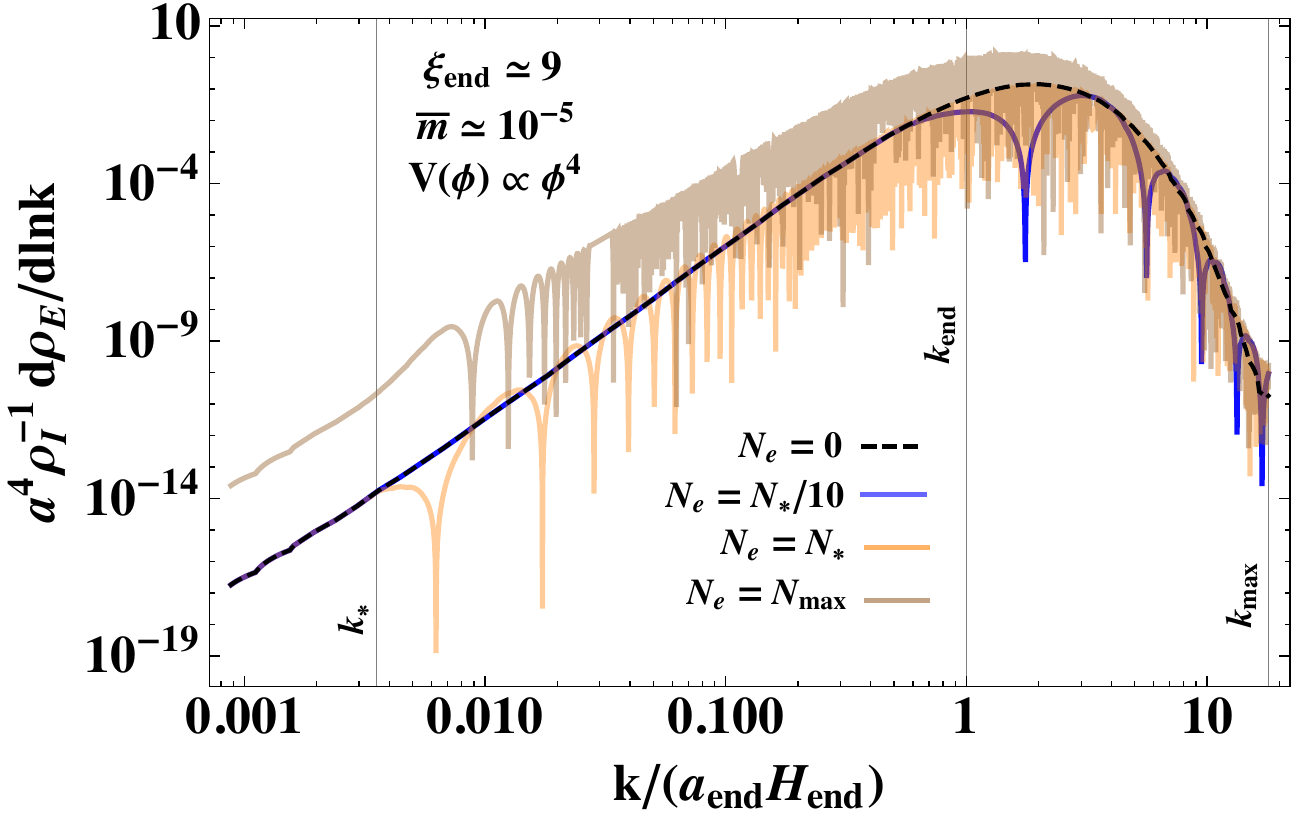}
\includegraphics[scale=.37]{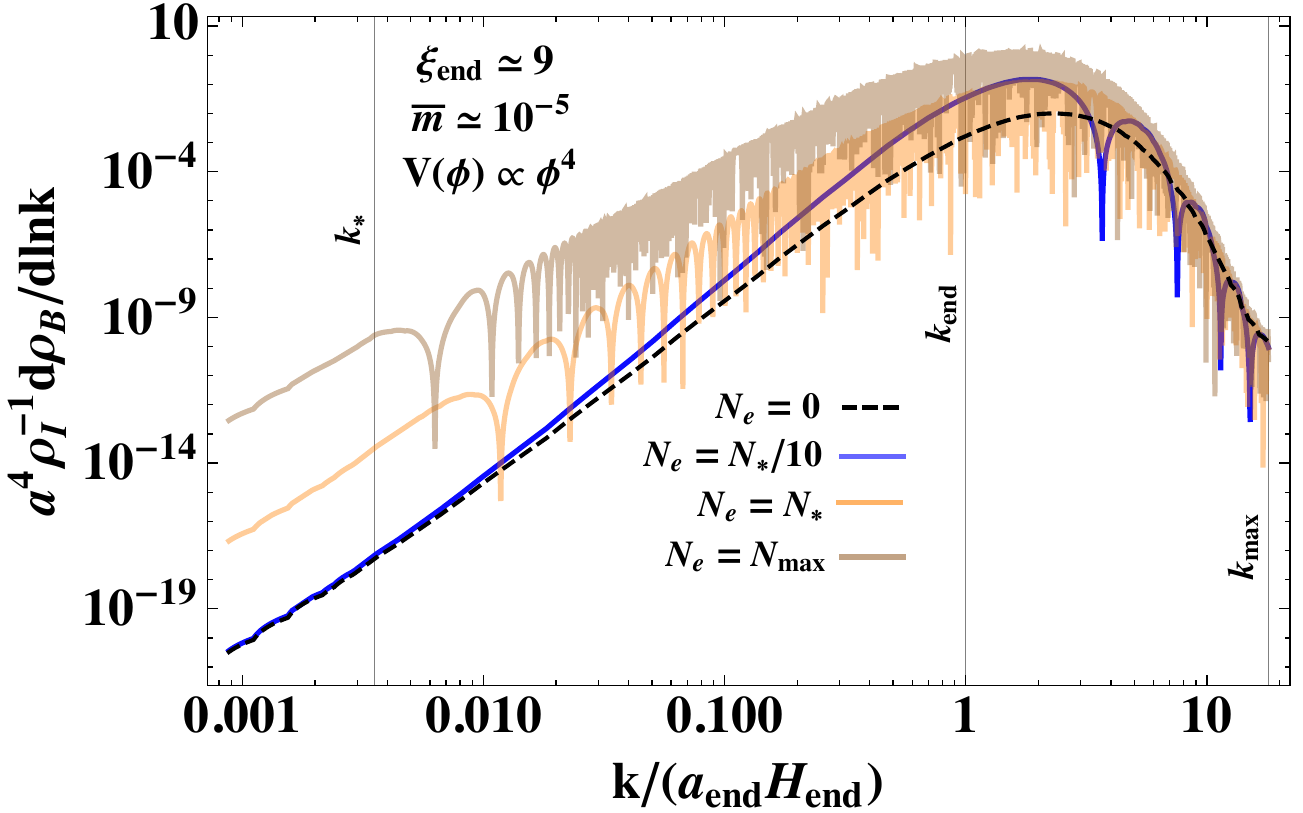}
\end{center}
\caption{On the left and right we show the electric and magnetic energy density spectra respectively (near the peak) for various numbers of e-folds after inflation, $N_e = 0\,({\rm dashed~black}), N_\ast/10\,({\rm blue}), N_\ast\,({\rm orange}), N_{\rm{max}}\,({\rm light~brown})$ where $a = \aend e^{N_e}$ defines the scale factor (see~\fref{confdiaglate}).~We have normalized the spectrum to the energy density of the inflaton at the end of inflation for $\xi_{\rm{end}} \simeq 9, \bar{m} \simeq 10^{-5}$ with a $\phi^4$ inflaton potential and factored out an $a^4$.}
\label{fig:lateenergyspec}
\end{figure}

In~\fref{lateenergyspec} and~\fref{lateenergyspec2} we summarize the evolution of the energy density spectrum after inflation focusing on modes with co-moving momentum $k_\ast \lesssim k \lesssim \kmax$ which contain the vast majority of the power.~In~\fref{lateenergyspec} on the left and right we show the electric and magnetic energy density spectra respectively while in~\fref{lateenergyspec2} we show their sum for various e-folds after inflation with $\xiend \simeq 9, \bar{m} \simeq 10^{-5}$ and a $V \propto \phi^4$ inflaton potential.~Note the spectra are similar for the non-minimal $\phi^4$ potential as seen in~\fref{energyspecsteep}.~The oscillatory behaviour seen in~\fref{lateenergyspec} arises from the oscillatory solutions for the modes (see~\eref{AIV} and~\eref{AV}) in regions IV and V of~\fref{confdiaglate}.~For modes $k_\ast < k < \kend$ this occurs once they re-enter the horizon while modes with $\kend < k < \kmax$, which contain the peak of the power spectrum, approach but never exit the horizon during inflation.~Modes with $k < k_\ast$ at $N > N_\ast$ remain outside the horizon for longer than modes with $k > \kast$ (see~\fref{confdiaglate}).~Thus they have not had enough time to begin oscillating upon reentering the horizon so we see a still smooth spectrum in this regime.~For the input values $\xiend =  9$ and $\bar{m} = 10^{-5}$ this gives in terms of e-folds $(N_e)$ after inflation $N_e = N_\ast \simeq 5.65,\,\bar{N} \simeq 11.30,\,N_{\rm{max}} \simeq 14.19$ where $a = \aend e^{N_e}$.~There are a number of features which are evident that reflect the behavior of the modes in the different limits of the equations of motion discussed in~\sref{evo} and can be understood with the help of~\fref{confdiaglate} and~\eref{scaleratios1},~\eref{scaleratios2}.
\begin{figure}[tbh]
\begin{center}
\includegraphics[scale=.55]{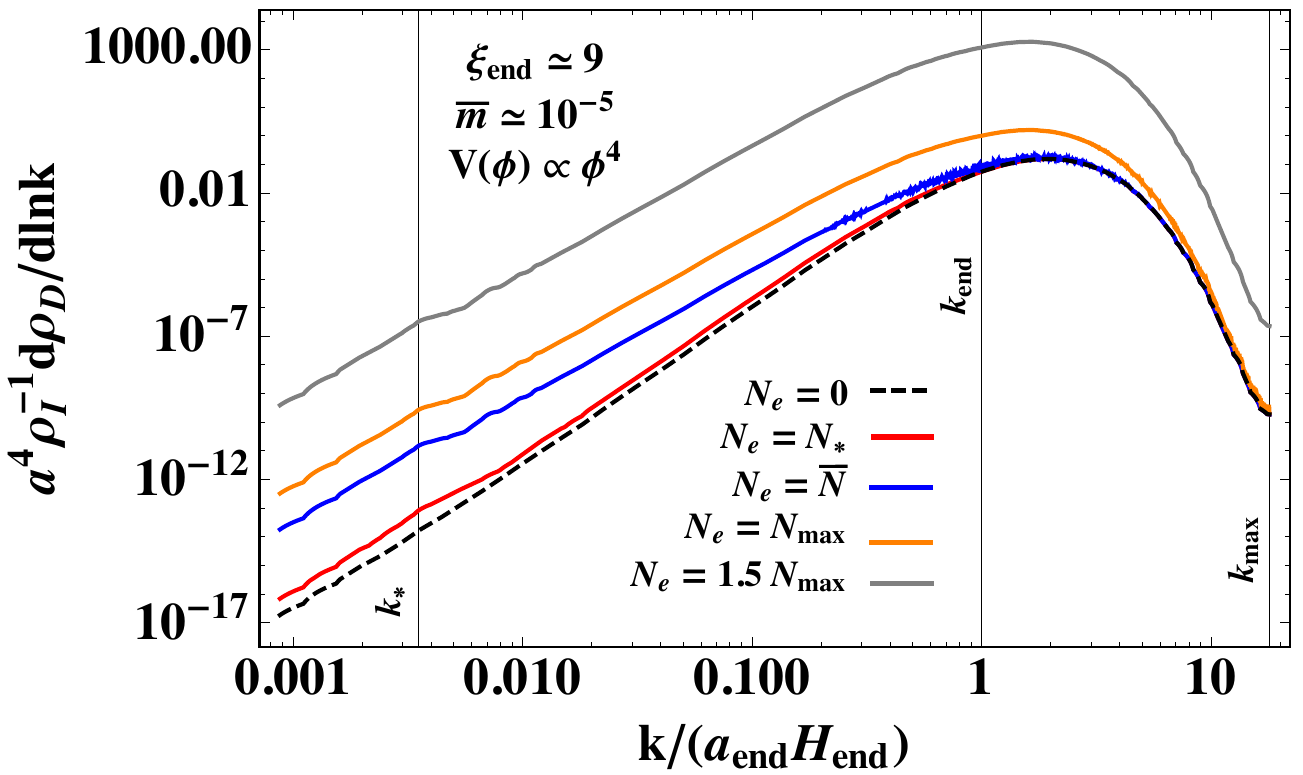}
\end{center}
\caption{Total energy density spectra (electric plus magnetic shown in~\fref{lateenergyspec})~for~$N_e = 0\,({\rm dashed~black}), N_\ast\,({\rm red}), \bar{N}\,({\rm blue}), N_{\rm{max}}\,({\rm orange}), 1.5\, N_{\rm{max}}\,({\rm grey})$ where $a = \aend e^{N_e}$ and again normalized to $\rhoIend$ with $a^4$ factored out.~See text for more information.}
\label{fig:lateenergyspec2}
\end{figure}

Looking first at the electric energy density spectrum on the left in~\fref{lateenergyspec}, we normalize to the energy density of the inflaton at the end of inflation and divide by $a^{-4}$ in order to compare to radiation like damping.~We see that once the input electric energy density spectrum is set at the end of inflation (black dashed) and which dominates over the magnetic component, the electric energy density then redshifts like radiation until $a_\ast$ (orange).~At this point scales larger than $k^{-1}_\ast$ begin redshifting like matter while modes at smaller scales around the peak continue redshifting like radiation.~Modes around the peak start becoming non-relativistic, and thus relatively enhanced as they begin redshifting like matter, once $q \leq m$ which happens at progressively later times for modes with larger and larger momenta as can be understood geometrically in~\fref{confdiaglate}.~The last mode to exit the horizon during inflation, $\kend = \aend\Hend$, becomes non-relativistic at $\bar{a}$ while the highest momentum tachyonic mode, $\kmax = 2\xi\,\kend$, becomes non-relativistic at $a_{\rm{max}}$ (light brown).~After this point, all of the modes redshift together like matter with $\propto a^{-3}$ damping and the shape of the electric energy density spectrum no longer changes.

On the right of~\fref{lateenergyspec} we examine the magnetic energy density spectrum which has a more interesting evolution.~We again normalize to the energy density of the inflaton at the end of inflation and factor out an $a^{4}$.~We see that after initially being sub dominant to the electric energy density at the end of inflation (black dashed), there is a brief period where modes around the peak damp like $\propto a^{-(2-3)}$ (blue) as compared to $\propto a^{-4}$ damping for the electric component.~As discussed in~\sref{evo}, this brief period of relative enhancement is due to the linear growth that the mode amplitudes in region II experience while they are super horizon (see~\eref{AII}) combined with continuity of the amplitude and derivative at the various boundaries.~After this brief period, the modes around the peak quickly begin redshifting like radiation with the usual $a^{-4}$ damping well before the time they reach $a_\ast$ (orange).~It is this very brief period of growth relative to matter, typically lasting around an e-fold or less after inflation, that allows the magnetic energy density to ``catch up" to the electric by the time $H = m$ at $a_\ast$ after which they redshift together like radiation and then eventually like matter (see~\fref{confdiaglate}).

In~\fref{lateenergyspec2} we show the total dark vector energy density spectrum given by the sum of the electric and magnetic components.~We see that when summed the oscillations of the electric and magnetic component cancel one another to give a smooth spectrum as can be shown analytically using~\eref{AIV} or~\eref{AV} and plugging it into~\eref{drhoEB}.~As we can also see, the total dark vector energy density spectrum essentially evolves like (the envelope of) the electric component.~This is because the electric component dominates at the end of inflation when the dark vector is produced and then redshifts along with the magnetic component once the magnetic component catches up as discussed above.~Thus to a good approximation the total energy density evolution has the same qualitative behavior as the electric component.~In particular, after inflation ends modes around the peak redshift like radiation until $q = m$ after which they redshift like matter (see~\fref{peakevo}) as the mass term dominates the dark vector energy density.

As we see from the energy density spectrum in~\fref{lateenergyspec}, the vast majority of the power is located at scales $\sim \kend^{-1}$ around the peak which are vastly smaller than the scales probed by CMB measurements.~To see this explicitly we write for modes around the peak $\sim\kend^{-1}$,
\bea\label{eq:kpeak}
1/k_{\rm end} &=& (\aend \Hend)^{-1} = (\aRH \epsilon_H H)^{-1} 
= (\frac{T_0}{\TRH} \epsilon_H H)^{-1} \nn
&\approx& 
\frac{10^{-1}}{T_0}\,\frac{ \epsilon_R}{ \epsilon_H} \left( \frac{\Mpl}{H} \right)^{1/2} 
\approx
10\ {\rm km} \ \frac{\epsilon_R}{\epsilon_H} \left( \frac{100 \ {\rm GeV}}{H} \right)^{1/2} \, ,
\eea
where we have assumed $\aend = \aRH$ and used~\eref{TRH} as well as $\aRH/a_0 = T_0/\TRH$ with the scale factor today set to $a_0 = 1$.~In the absence of an extreme hierarchy between $\epsilon_R$ and $\epsilon_H$, we see that the typical co-moving scale associated with the peak is $\lesssim 10\,\rm{km}$ which is tiny on cosmological scales and thus we expect isocurvature to be negligible on the large scales relevant for the CMB.~However, since it is relevant for how the dark matter is distributed spatially after matter radiation equality and the evolution of density perturbations, below we compute the power spectrum of isocurvature perturbations.~If the vector has a mass already during inflation, there could be regions of parameter space where the longitudinal and transverse components give comparable contributions to the dark matter energy density.~As discussed below, in this case there would be a double peaked structure in the energy density spectrum with one peak corresponding to the transverse component located at co-moving momenta $\sim k_{\rm{end}}$ and a second one corresponding to the longitudinal mode at $k_\ast$~\cite{Graham:2015rva}.

%

\subsection{Isocurvature and density contrast power spectrum} \label{sec:isopec} 

Up until now, we have considered only the mean energy density and spectra which can be written in terms of the dark vector field power spectrum as given in~\eref{rhoDint}.~However, the dark vector energy density is subject to fluctuations which can be of the same order as the mean energy density.~These fluctuations will be independent of the inflaton ones which set the curvature (or adiabatic) perturbations that are imprinted on the metric.~The fluctuations in the dark vector energy density will therefore contribute to isocurvature perturbations and can have implications for the CMB as well as structure formation and clumping~\cite{Kolb:1994fi} of the dark matter.~Since measurements of the CMB~\cite{Akrami:2018odb} severely constrain the amplitude of isocurvature perturbations, we must ensure that they are suppressed on long length scales.~Following~\cite{Graham:2015rva,Alonso-Alvarez:2018tus}, here we compute these isocurvature perturbations and demonstrate explicitly that they are highly suppressed at the large scales relevant for the CMB.

The starting point is the density contrast field $\delta(\vec x)$ which describes deviations from the mean dark vector energy density $\langle \rho \rangle$ and is defined as,
\bea \label{rhoqu}
\rho(\vec x) = \langle \rho \rangle (1 + \delta(\vec x)) .
\eea
Near the time of matter radiation equality and once the vector dark matter begins redshifting like matter, we can describe the energy density via the mass term in the magnetic component.~As we derive in the Appendix, the Fourier transform of $\delta(\vec x)$ can be written in terms of products of the power spectrum for the tachyonically enhanced transverse mode as,
\bea\label{eq:deltaPS}
{\cal P}_{\delta}(k,t) & = 
\frac{k^2}{\left[ \int_0^\infty \frac{dk'}{k'} {\cal P}_{A_+} (k',t)  \right]^2}
\int_0^\infty dq \int_{|q-k|<p<q+k}  dp \ 
\frac{1}{q^2 p^2}  {\cal P}_{A_+}(p,t) {\cal P}_{A_+}(q,t)\, ,
\eea
where we have taken the mass term in~\eref{rhoDint} to represent the energy density at late times and defined the power spectrum in terms of the 2-point function of a random variable $X$,
\bea \label{powdqu}
\langle X(\vec k) X(\vec k') \rangle = 
(2\pi)^3 \delta^3(\vec k + \vec k') \frac{2\pi^2}{k^3} {\cal P}_{X}(k) \, ,
~~~X \equiv A_+, \delta.
\eea
We see in~\eref{deltaPS} that the power spectrum for $\delta$ corresponding to the transverse vector mode\,\footnote{We note that~\eref{deltaPS} agrees with the result found in~\cite{Alonso-Alvarez:2018tus} for \emph{scalar} dark matter produced during inflation.} differs from the one corresponding to the longitudinal mode~\cite{Graham:2015rva}.

Performing the integral in~\eref{deltaPS} numerically we show the density contrast power spectrum in~\fref{density} for $\bar{m} \ll 1$ and $\xiend \approx 9$.~We see explicitly that the spectrum falls off sharply at large scales\footnote{Note the slope at large scales is universal as the underlying energy density spectrum in general falls off like $k^4$ (see Appendix B) while the convolution integral in~\eref{deltaPS} modifies this to $\sim k^3$ for the density contrast power spectrum.} like $\sim k^3$.~We also see that the location of the peak does not shift noticeably from the peak in the mean energy density spectrum.~At scales relevant for the CMB we see the power is completely negligible though for $m \sim \Hend,\xi_{\rm{CMB}} \sim 2.5$ perhaps these isocurvature perturbations may have observable consequences, but we do not explore this possibility here.~To determine the matter distribution and scale of clumping of the vector dark matter today, one needs to follow the evolution of these density perturbations through the non-linear regime until today.~The power spectrum in~\fref{density} serves as the input for this evolution~\cite{Berges:2019dgr}, beginning at matter radiation equality, but further investigation is left to future work. 
\begin{figure}[tbh]
\begin{center}
\includegraphics[scale=.5]{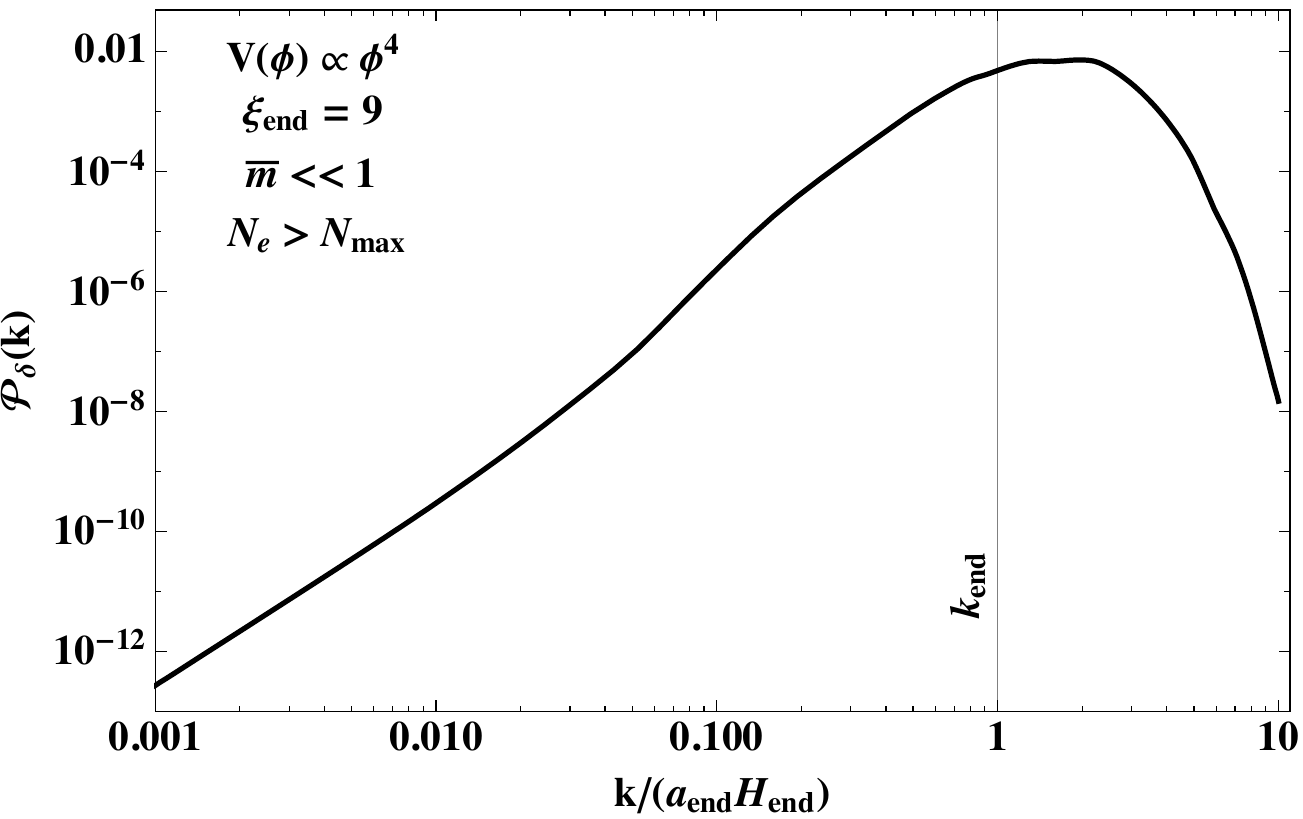}
\end{center}
\caption{The late time density contrast power spectrum for $\bar{m} \ll 1$ and $\xiend \approx 9$.}
\label{fig:density}
\end{figure}

Finally, we comment that in the mechanism presented here, isocurvature for the transverse mode is suppressed due to the fact that the tachyonic modes experience maximal growth when they are of order the horizon combined with the time dependence of the tachyonic instability parameter $\xi$ \emph{during} inflation.~This is in contrast to the mechanism for suppressing isocurvature in the energy density of the longitudinal mode\,\footnote{Note this is also distinct from how isocurvature is suppressed on long length scales for scalar dark matter production mechanisms~\cite{Alonso-Alvarez:2018tus,Berges:2019dgr,Markkanen:2018gcw} which are also connected to inflation.} which is due to how the different modes redshift \emph{after} inflation~\cite{Graham:2015rva}.

\subsection{Clumping of vector dark matter} \label{sec:clumping} 

As discussed in other `clumpy' dark matter scenarios~\cite{Graham:2015rva,Alonso-Alvarez:2018tus,Berges:2019dgr}, the scale corresponding to the peak of the power spectrum of energy density fluctuations also has implications for the scale on which the dark matter `clumps'.~In one of these scenarios, a dark vector has a mass during inflation and the longitudinal mode is necessarily produced by inflationary fluctuations~\cite{Graham:2015rva}.~In this case, due to how the modes redshift after inflation, a peak is produced in the energy density spectrum at $k_\ast$ (see~\fref{confdiaglate}) where,
\bea
1/k_\ast \sim 10^{10} \,{\rm km} \times 
\sqrt{ \frac{10^{-5} \, {\rm eV }}{m} } .
\eea
We see that for the longitudinal mode the location of the peak only depends on the dark vector mass which is constrained to be $m \geq 10^{-5} \, {\rm eV }$ if it makes up the entirety of the dark matter~\cite{Graham:2015rva}.~Considering the range of Hubble scales during inflation $10^2 < H/{\rm GeV} < 10^{14}$ this leads to the range of `clumping' scales for the longitudinal component,  
\bea
10^{-1} \, {\rm km} \ < k_\ast^{-1} < 10^{10} \, {\rm km} \, .
\eea

In the case of the transverse component produced via tachyonic instability, we saw in~\eref{kpeak} that the location of the peak does not depend on the dark vector mass, but instead on the Hubble scale during inflation as well as $\epsilon_R$ and $\epsilon_H$.~For the same range of Hubble scales during inflation $10^2 < H/{\rm GeV} < 10^{14}$ and taking $\epsilon_H / \epsilon_R \sim 1$ we find the transverse component instead clumps on scales in the range,
\bea
{\rm cm} < k_{\rm end}^{-1} 
< 10 \, {\rm km} .
\eea
The locations of the two peaks are related by the Hubble scale and dark vector mass,
\bea
\frac{k_{\rm end}^{-1} }{k_*^{-1} }= 
\frac{\epsilon_R}{\epsilon_H}
\left( \frac{m}{H} \right)^{1/2} 
\sim \left( \frac{m}{H} \right)^{1/2} \, .
\eea
We see in general the transverse mode clumps on much smaller scales than the longitudinal mode though when $m\sim H$ the scales can be comparable.~As discussed, this also opens the possibility of a double peak in the power spectrum when both the longitudinal and transverse components contribute appreciably to the relic abundance and would imply dark matter clumping on two different scales.~This `doubly clumpy' possibility can of course only occur if the dark vector has a mass already during inflation and thus would constitute a striking signal of vector dark matter with an inflationary origin.

\section{{\bf Summary and Outlook}}

In this study we have examined in detail the recently proposed mechanism~\cite{Bastero-Gil:2018uel} for producing non-thermal dark photon dark matter at the end of inflation.~This mechanism can generate the observed dark matter relic abundance for dark vector masses in the range $\mu\,{\rm eV} \lesssim m \lesssim 10\,{\rm TeV}$ and Hubble scales during inflation in the range $100\,{\rm GeV} \lesssim H \lesssim 10^{14}\,{\rm GeV}$.~We have focused in particular on the case where the dark vectors are relativistic at the time their mass is generated and examined the associated cosmic evolution to compute the relic abundance today.~We have also examined the power spectrum and cosmic evolution of the dark vector modes demonstrating explicitly that the late time spectrum preserves the peak generated at the end of inflation.~We have shown that the peak corresponds to small physical scales today, $\ell_{\rm today} \sim {\rm cm} -  100\,{\rm km}$, with large density fluctuations at $\ell_{\rm today}$ implying a clumpy nature for the vector dark matter.~The case of a non-relativistic dark vector at the time its mass is generated has been left to forthcoming work.

There are a number of interesting avenues to explore the phenomenology associated with the dark photon dark matter production mechanism presented here.~If there are other dark charged particles present during inflation they can potentially be produced via a dark Schwinger production mechanism.~In this case dark charged fermions and scalars may also contribute to the final dark matter relic abundance.~If the dark vector obtains its mass via a dark Higgs mechanism, this can lead to an alternative cosmic evolution to the one explored in this study and there may also be a possibility of generating gravitational wave signals associated with the dark Higgs phase transition~\cite{Breitbach:2018ddu}.~Allowing for non-zero (but small) kinetic mixing between the dark and visible photons can lead to interesting dark matter phenomenology as has been thoroughly explored in many studies.~Whether any of the initial polarization of the dark vector survives cosmic evolution and has observable effects is also worth investigating.~Explorations of these various interesting possibilities are ongoing.

\vspace{0.5cm}
\noindent
{\bf \emph{Acknowledgements}}:
We thank Prateek Agrawal, Diego Blas, Adam Falkowski, Bohdan Grzadkowski, Takeshi Kobayashi, Eric Madge, Manuel Masip, Gilad Perez, Maxim Pospelov, Jennifer Schober, Pedro Schwaller, Javi Serra, Anna Socha, Walter Tangarife, Tomer Volansky, and Tien-Tien Yu for useful comments and discussions.~This work has been partially supported by MINECO grants PID2019-106087GB-C22,~including ERDF (J.S.,\,R.V.M.),~PID2019-105943GB-I00 (M.B.G.),~Junta de Andaluc\'{i}a Projects FQM-101, A-FQM-211-UGR18, P18-FR-4314, (fondos FEDER), and SOMM17/6104/UGR (M.B.G.,\,J.S.,\,
R.V.M.).~L.U.~acknowledges support from the PRIN project ``Search for the Fundamental Laws and Constituents'' (2015P5SBHT\textunderscore 002).~R.V.M. would also like to acknowledge the Mainz Institute for Theoretical Physics (MITP) of the Cluster of Excellence PRISMA+ (Project ID 39083149) for their hospitality and partial support as well as participants of \emph{The Mysterious Universe} workshop for useful discussions and stimulating atmosphere while part of this work was completed.

\section*{{\bf Appendix}}\label{sec:app}

\subsection{Derivation of equations of motion, energy density, and pressure}\label{sec:appA}

Here we derive the equations of motion as well as the energy and pressure densities.

\begin{center}
{\bf\emph{Conventions}}
\end{center}

Before presenting the derivation we define the conventions used.~For the metric we have,
\bea
ds^2 = -dt^2 + a^2(t) d\vec x^2 = a^2(\tau)(-d\tau^2 + d\vec x^2) \, ,
\eea
\bea
g_{\mu\nu} &=& 
\begin{pmatrix}
-1 & 0 \\
0 & a^2(t) \delta_{ij}
\end{pmatrix} =
a^2(\tau)
\begin{pmatrix}
-1 & 0 \\
0 &  \delta_{ij}
\end{pmatrix} \, , \nn
g^{\mu\nu} &=& 
\begin{pmatrix}
-1 & 0 \\
0 & a^{-2}(t) \delta^{ij}
\end{pmatrix}=
a^{-2}(\tau)
\begin{pmatrix}
-1 & 0 \\
0 & \delta^{ij}
\end{pmatrix} \,, ~~~~~~
\eea
with $t$ the cosmic time and $\tau$ the conformal time.~The Levi-Civita tensor is given by,
\bea
\epsilon^{\mu\nu\rho\sigma} = \frac{\tilde \epsilon^{\mu\nu\rho\sigma}}{\sqrt{-g}} \, , \quad 
\epsilon_{\mu\nu\rho\sigma} = \sqrt{-g} \ \tilde \epsilon_{\mu\nu\rho\sigma} \, ,
\eea
with the following conventions for the anti-symmetric Levi-Civita symbol and metric,
\bea
\tilde \epsilon^{0123} = + 1\, , \quad \tilde\epsilon_{0123} = -1 \, ,\nn
\sqrt{-g} \equiv \sqrt{- \det (g_{\mu\nu})} = a^3(t) = a^4(\tau) \, .
\eea
The three-dimensional Levi-Civita symbol is related to the four-dimensional one as
\bea
\tilde \epsilon_{ijk} = \tilde \epsilon^{ijk} = \tilde \epsilon^{0ijk} \, .
\eea
For the Hubble parameter we have in terms of the scale factors,
\bea
H \equiv \frac{1}{a(t)} \frac{da(t)}{dt} = \frac{\dot a(t)}{a(t)} \, , \qquad {\cal H} \equiv \frac{1}{a(\tau)} \frac{da(\tau)}{d\tau} = \frac{a'(\tau)}{a(\tau)} \, ,
\eea
where we use an overdot for the cosmic time derivative and a prime for the conformal time derivative.~We work in comoving momentum space where for the classical field $A_\mu$ we have:
\begin{align}
A_0(\vec x, \tau) & = \int \frac{d^3 k}{(2\pi)^3} A_0(\vec k, \tau) e^{i \vec k \cdot \vec x} \, , \\
A_i(\vec x, \tau) & = \int \frac{d^3 k}{(2\pi)^3} A_i(\vec k, \tau) e^{i \vec k \cdot \vec x} \, .
\end{align} 
As the gauge field is real, $A_\mu(\vec x, \tau) = A^*_\mu(\vec x, \tau)$ which implies $A^*_\mu(\vec k, \tau) = A_\mu(-\vec k, \tau)$.~We decompose $\vec A(\vec k,\tau)$ along its transverse ($\vec A_T$) and longitudinal components ($A_L$),
\begin{align}
\vec k \cdot \vec A_T & = 0 \\
\vec k \cdot \vec A & = |\vec k| A_L \equiv k A_L \, ,
\end{align}
and further decompose the transverse modes into the usual two polarizations
\be \label{eq:trpola}
\vec A_T(\vec k,\tau) = \sum_{\lambda = \pm} \vec \epsilon_\lambda (\vec k) \left[ A_\lambda(k,\tau) + A_\lambda(k,\tau)^* \right]  \, ,
\ee
where the polarization vectors satisfy,
\begin{align}
& \vec k \cdot \vec \epsilon_\pm(\vec k) = 0 \, , \quad \vec k \times \vec \epsilon_\pm(\vec k) = \mp i  k \vec \epsilon_\pm(\vec k)\, , \\ 
&  \vec k \cdot \vec \epsilon_L (\vec k) = k \, , \quad \vec k \times \vec \epsilon_L(\vec k) = 0 \, , \\
& \vec \epsilon_\lambda(\vec k)^* = \vec \epsilon_\lambda(-\vec k)\, , \quad 
\vec \epsilon_\lambda(\vec k) \cdot \vec \epsilon_{\lambda'}(- \vec k) = \delta_{\lambda \lambda'} \, .
\end{align}
The power spectra associated to the two-point correlation functions of the classical field is,
\begin{align}
\langle \vec A(\vec k,t) \cdot \vec A(\vec k',t) \rangle & = (2\pi)^3 \delta^3(\vec k + \vec k') \frac{2\pi^2}{k^3} {\cal P}_A(k,t) \, , \nonumber \\
\langle \partial_t \vec A(\vec k,t) \cdot \partial_t \vec A(\vec k',t) \rangle & = (2\pi)^3 \delta^3(\vec k + \vec k') \frac{2\pi^2}{k^3} {\cal P}_{\partial_t A}(k,t) \, ,
 \label{eq:powspclass}
\end{align}
where we can then write for the two point function in position space,
\bea \label{eq:powspdef}
\langle \vec A(\vec x,t)^2 \rangle = \int_0^\infty \frac{dk}{k} {\cal P}_A(k,t) = \int \frac{d^3k}{(2\pi)^3} \frac{2\pi^2}{k^3} {\cal P}_A(k,t) \, .
\eea

For the quantum field we expand in terms of creation and annihilation operators,
\bea \label{eq:Aqu}
\hat{\vec A} (\vec x,t) = \sum_{\lambda = \pm, L} \int \frac{d^3k}{(2\pi)^3} e^{i \vec k \cdot \vec x} \ \vec \epsilon_\lambda(\vec k) \ [A_\lambda(k,t) a_\lambda(\vec k) + A_\lambda(k,t)^* a_\lambda^\dagger (-\vec k) ] \, .
\eea
where the creation and annihilation operators satisfy,
\bea
a_\lambda(\vec k) |0\rangle = 0 \, , \quad \langle 0| a_\lambda^\dagger (\vec k) = 0\, , \quad 
\left[a_\lambda(\vec k), a^\dagger_{\lambda'}(\vec k')  \right] = (2\pi)^3 \delta_{\lambda \lambda'} \delta^3(\vec k - \vec k') \, .
\eea
Two point correlation functions with the quantum field are then obtained by sandwiching the field operators between vacuum states,
\bea\label{eq:A2quantum}
\langle \hat{\vec A}(\vec x,t)^2 \rangle &\equiv& 
\langle 0| \hat{\vec A}(\vec x,t)^2  |0\rangle \nn
&=& \sum_{\lambda,\lambda'} \int \frac{d^3k}{(2\pi)^3} \frac{d^3k'}{(2\pi)^3} 
e^{i (\vec k +\vec k') \cdot \vec x}
\vec \epsilon_\lambda(\vec k) \cdot \vec \epsilon_{\lambda'}(\vec k') \nn
&\times&\langle 0| (A_\lambda(k,t) a_\lambda(\vec k) A_{\lambda'}(k',t)^* a^\dagger_{\lambda'}(-\vec k')) |0 \rangle  \nn 
 &=& \sum_\lambda \int\frac{d^3 k }{(2\pi)^3} |A_\lambda(k,t)|^2 \, ,~~~~
\eea
where to go from second to third line we used the commutation relations.~Comparing to the definition of the power spectrum in~\eref{powspdef} we find in the quantum case,
\bea \label{eq:powquantum}
{\cal P}_{\hat A}(k,t) = \frac{k^3}{2\pi^2} \sum_\lambda |A_\lambda(k,t)|^2 \, .
\eea
Similarly we can compute the power spectrum for the field time derivative $\langle (\partial_t\hat{\vec A}(\vec x,t))^2 \rangle$,
\bea
\langle (\partial_t\hat{\vec A}(\vec x,t))^2 \rangle &=&  
\sum_\lambda \int\frac{d^3 k }{(2\pi)^3} |\partial_t A_\lambda(k,t)|^2 \,, \nn
{\cal P}_{\partial_t \hat A}(k,t) &=& 
\frac{k^3}{2\pi^2} \sum_\lambda |\partial_t A_\lambda(k,t)|^2 \, .
\eea
Note that the time dependence in the Fourier expansion is only in the function $A_\lambda(k,t)$.

\begin{center}
{\bf\emph{Equations of motion}}
\end{center}

To derive the equations of motion we consider the following action,
\bea\label{eq:Lagrangian}
S = \int d^4 x \, L &=& \int d^4 x  
\sqrt{-g} \Big[ -\frac{1}{2}g^{\mu\nu} \partial_\mu \phi \partial_\nu \phi - V(\phi) -\frac{1}{4} g^{\mu\nu}g^{\rho\sigma}F_{\mu\rho}F_{\nu\sigma} \nn
&-& \frac{\alpha}{4f} \phi \frac{1}{2} \epsilon^{\mu\nu\rho\sigma} F_{\mu\nu}F_{\rho\sigma} - \frac{1}{2} m^2 g^{\mu\nu}A_\mu A_\nu \Big] \, , 
\eea
where the field strength is defined in the usual way as,
\be
F_{\mu\nu} = \partial_\mu A_\nu - \partial_\nu A_\mu \, .
\ee
Note that we have inserted the mass term for the vector field \`a la Proca so this Lagrangian describes a model with a scalar field and a massive vector field and is strictly speaking not a gauge theory, but this distinction is not important for present purposes.~Note also that,
\bea
\frac{1}{2} \epsilon^{\mu\nu\rho\sigma}F_{\mu\nu} F_{\rho\sigma} &=& 
\frac{4}{\sqrt{-g}} \left[ (\partial_\tau \vec A - \nabla A_0) \cdot (\nabla \times \vec A) \right]  \\
&\rightarrow&  \frac{4}{\sqrt{-g}} \int \frac{d^3 k \ d^3 k'}{(2\pi)^6} e^{i(\vec k + \vec k') \cdot \vec x}  (\partial_\tau \vec A(\vec k,\tau) - i \vec k A_0(\vec k, \tau)) \cdot (i \vec k'  \times \vec A(\vec k', \tau))  \, ,\nonumber
\eea
where we see that this vanishes for the longitudinal mode for which $\vec k'  \times \vec A(\vec k', \tau) = 0$.~Thus, the term with $F \tilde F$ only affects the transverse modes as we will see explicitly below.

The equations of motion are obtained from the lagrangian density via,
\begin{align}
\partial_\alpha \frac{\delta L}{\delta(\partial_\alpha \phi)} - \frac{\delta L}{\delta \phi} & = 0 \, , \label{eq:EOMphiAP} \\
\partial_\alpha \frac{\delta L}{\delta(\partial_\alpha A_\beta)} - \frac{\delta L}{\delta A_\beta} & = 0  \label{eq:EOMA} \, .
\end{align}
From the first we equation obtain in cosmic time,
\bea
\ddot \phi - \frac{1}{a^2(t)}\partial_i^2 \phi + 3 H \dot \phi + \frac{\partial V}{\partial\phi} + \frac{\alpha}{4 f} \frac{1}{2} \epsilon^{\mu\nu\rho\sigma} F_{\mu\nu}F_{\rho\sigma} = 0 \, ,
\eea
or in conformal time,
\bea
\partial_\tau^2 \phi - \partial_i^2 \phi + 2 {\cal H} \partial_\tau \phi + a^2(\tau) \frac{\partial V}{\partial\phi} + a^2(\tau) \frac{\alpha}{4 f} \frac{1}{2} \epsilon^{\mu\nu\rho\sigma} F_{\mu\nu}F_{\rho\sigma} = 0 \, .
\eea

Next, we turn to the equations of motion for the vector field which we derive in conformal time.~From here on every time we write $a$ for the scale factor we mean implicitly $a(\tau)$.~After some algebra~\eref{EOMA} can be cast into the form,
\bea
\sqrt{-g} \left[ g^{\alpha\nu} g^{\beta\sigma} \partial_\alpha F_{\nu\sigma}  
 +\frac{\alpha}{2f} (\partial_\alpha \phi) \epsilon^{\alpha\beta \rho \sigma}F_{\rho\sigma}  - m^2 g^{\beta \mu} A_\mu \right] & = 0 \, , \nonumber \\
 \eta^{\alpha\nu} \eta^{\beta\sigma} \partial_\alpha F_{\nu\sigma}  
 +\frac{\alpha}{2f} (\partial_\alpha \phi) \tilde\epsilon^{\alpha\beta \rho \sigma}F_{\rho\sigma}  - a^2 m^2 \eta^{\beta \mu} A_\mu & = 0 \, ,
 \label{eq:EOMAbeta}
\eea
where in the second line we have the flat metric $\eta^{\mu\nu} = \eta_{\mu\nu} = (-,+,+,+)$.
This gives four equations of motion for the massive vector field where the $\beta = 0$ and $\beta = l$ components can be written respectively as,
\begin{align} \label{eq:beta0}
& (\partial_i^2 A_0 - \partial_0 \partial_i A_i ) + \frac{\alpha}{f} \tilde \epsilon_{ijk} (\partial_i \phi) \partial_j A_k  - a^2 m^2  A_0 = 0 \,,\\ 
& \partial_0 \partial_0 A_l - \partial_i \partial_i A_l - \partial_l \partial_0 A_0 + \partial_l \partial_i A_i 
 - \frac{\alpha}{f} \tilde\epsilon_{ljk} \left[ (\partial_0\phi)  \partial_j A_k  +  (\partial_j \phi) (\partial_k A_0 - \partial_0 A_k)  \right] + a^2 m^2 A_l = 0 \, . 
 \label{eq:betal}
\end{align}
%
As a massive vector has three degrees of freedom, we need to apply one constraint which can be obtained by acting on \eref{EOMAbeta} with $\partial_\beta$,
\bea \label{eq:Procalikeconst}
\partial_\beta \left[   \eta^{\alpha\nu} \eta^{\beta\sigma} \partial_\alpha F_{\nu\sigma}  
 +\frac{\alpha}{2f} (\partial_\alpha \phi) \tilde\epsilon^{\alpha\beta \rho \sigma}F_{\rho\sigma}  - a^2 m^2 \eta^{\beta \mu} A_\mu \right] = 0 \, .
\eea
The only terms that does not vanish by symmetry is the last, which gives
\be \label{eq:Procafull}
2{\cal H} A_0 + \partial_0 A_0 -  \partial_i A_i = 0 \, .
\ee
Plugging this constraint into~\eref{beta0} leads to
\bea
\partial_\tau^2 A_0 + 2 {\cal H} \partial_\tau A_0 + (2 \partial_\tau {\cal H} + a^2 m^2 - \nabla^2)A_0 = \frac{\alpha}{f} \nabla \phi \cdot (\nabla \times \vec A) \, ,
\label{eq:condA0}
\eea
while using \eref{Procafull} in \eref{betal} we get
\bea\label{eq:EOAfull}
(\partial_\tau^2 - \nabla^2 + a^2 m^2)\vec A + 2 {\cal H} \nabla A_0 = \frac{\alpha}{f} \left[ \partial_\tau \phi (\nabla \times \vec A) - \nabla\phi \times (\partial_\tau \vec A - \nabla A_0)  \right] \, .
\eea

We now proceed by first projecting along the longitudinal mode and then along the transverse modes.

For the longitudinal mode we have,
\bea
\tilde \epsilon_{ijk} (\partial_i \phi) \partial_j A_k \rightarrow (\nabla \phi(\vec x, \tau)) \cdot (i \vec k \times \vec A(\vec k, \tau)) = 0\, .
\eea
This then gives for~\eref{beta0},
\bea
(\partial_i^2 A_0 - \partial_0 \partial_i A_i )   - a^2 m^2  A_0 = 0 \, ,
\eea
which in Fourier space has the solution,
\bea \label{eq:A0}
A_0(\vec k,\tau) = \frac{-i  \vec k \cdot \partial_\tau \vec A(\vec k, \tau)}{k^2 + a^2 m^2} \, . 
\eea
We next consider the terms in squared parentheses in~\eref{betal}, 
\begin{align}
& \tilde\epsilon_{ljk} \left[ (\partial_0\phi)  \partial_j A_k  +(\partial_j \phi) (\partial_k A_0 - \partial_0 A_k)  \right] \nonumber \\
& \rightarrow  (\partial_\tau \phi) \left( i \vec k \times \vec A_L(\vec k , \tau) \right) + (\nabla \phi) \times \left( i \vec k A_0(\vec k, \tau) - \partial_\tau \vec A_L(\vec k, \tau) \right) \, ,
\end{align}
This results in vectors which are orthogonal to the direction of $\vec k$ and so do not contribute to the equation of the longitudinal mode.~In Fourier space~\eref{betal} then simplifies to,
\bea
\partial_\tau^2 A_L(\vec k,\tau) - i k \partial_\tau A_0(\vec k,\tau) + a^2 m^2 A_L(\vec k,\tau) = 0 \, .
\eea
Using the solution for $A_0$ in~\eref{A0} this becomes,
\bea \label{eq:ALtau}
\partial_\tau^2 A_L(\vec k,\tau) + \frac{2k^2}{k^2 + a^2 m^2} {\cal H} \partial_\tau A_L(\vec k,\tau) + (k^2 + a^2 m^2) A_L(\vec k,\tau) = 0 \, . 
\eea
Switching to cosmic time the equation of motion for the longitudinal mode becomes,
\bea \label{ALt}
\ddot A_L(\vec k,t) + \frac{3k^2 + a^2(t) m^2}{k^2 + a^2(t) m^2} H \dot A_L(\vec k,t) + \left( \frac{k^2}{a^2(t)} + m^2 \right) A_L(\vec k,t) = 0 \, ,
\eea
which is in agreement with~\cite{Graham:2015rva}.

Turning to the transverse modes, for which $\vec k \cdot \vec A = 0$, from~\eref{EOAfull} we must drop the term 
$2{\cal H} \nabla A_0 = 2 i {\cal H} \vec k A_0$, which is along the vector $\vec k$, so we have
\bea\label{eq:EOMAT}
(\partial_\tau^2 - \nabla^2 + a^2 m^2)\vec A_T  = \frac{\alpha}{f} \left[ \partial_\tau \phi (\nabla \times \vec A_T) - \nabla\phi \times (\partial_\tau \vec A_T - \nabla A_0)  \right]
\eea
In cosmic time this becomes
\bea \label{eq:comparePrateek}
\ddot{\vec A}_T + H \dot{\vec A}_T - \frac{\nabla^2}{a^2} \vec A_T + m^2 \vec A_T - \frac{1}{a} \frac{\alpha}{f} \left[ \dot\phi (\nabla \times \vec A_T) -  \nabla \phi \times \dot{\vec A}_T + \frac{1}{a} \nabla\phi \times \nabla A_0 \right] = 0 \, .
\eea

Next we consider the case of a homogeneous scalar field, $\phi(\vec x,t) \approx \phi(t)$, as appropriate for $\phi$ during inflation, and we approximate $\nabla \phi \approx 0$ .~Using~\eref{trpola} for the transverse modes, \eref{EOMAT} in Fourier space simplifies to,
\bea
\partial_\tau^2 A_{\pm}(k,\tau) + \left[ k^2 \mp \frac{\alpha}{f}(\partial_\tau\phi) k + a^2 m^2 \right] A_{\pm}(k,\tau) = 0 \, .
\eea
Using the fact that the conformal time during inflation is $\tau \approx -(aH)^{-1}$ we can rewrite, 
\bea
\partial_\tau \phi = a \frac{\partial\phi}{\partial t} = aH \frac{\dot\phi}{H} \approx - \frac{1}{\tau} \frac{\dot\phi}{H} \, .
\eea
The equation of motion then reads in conformal time,
\bea \label{eq:EOM3}
\partial_\tau^2 A_{\pm}(k,\tau) + \left[ k^2 \pm \frac{\alpha}{f}\frac{\dot\phi}{H}\frac{k}{\tau} + a^2 m^2 \right] A_{\pm}(k,\tau) = 0 \, .
\eea
Using $\partial_\tau^2 A = a^2(t) H \dot A + a^2(t) \ddot A$ this can be rewritten in cosmic time as,
\bea \label{eq:Apmt}
\ddot A_{\pm}(k,t) + H \dot A_{\pm}(k,t) + \left[ \frac{k^2}{a^2(t)} \mp \frac{\alpha}{f} \dot\phi \frac{k}{a(t)} + m^2 \right] A_{\pm}(k,t) = 0 \, .
\eea


\vspace{2mm}
\begin{center}
{\bf \emph{Energy and pressure densities}}
\end{center}

Starting from the action, we can compute the stress-energy tensor via,
\bea
T_{\alpha\beta} = -\frac{2}{\sqrt{-g}} \frac{\delta S}{\delta g^{\alpha\beta}} \, ,
\eea
where the following relations will be useful for the calculation,
\bea
\frac{\delta}{\delta g^{\alpha\beta}} \sqrt{-g} = -\frac{1}{2} \sqrt{-g} \ g_{\alpha\beta} \, , \qquad \frac{\delta}{\delta g^{\alpha\beta}}(-g) = g \ g_{\alpha\beta} \, .
\eea
One can verify that after computing $\frac{\delta S}{\delta g^{\alpha\beta}}$, there are two terms proportional to $\phi F \tilde F$ which cancel exactly.~Thus, we see that the operator responsible for inducing the tachyonic instability does not contribute to the energy density.~Explicitly we find,
\bea
T_{\alpha\beta} &=&  \partial_\alpha \phi \partial_\beta \phi + g^{\mu\nu} F_{\mu\alpha}F_{\nu\beta} + m^2 A_\alpha A_\beta \nn
&-& g_{\alpha\beta} \left( \frac{1}{2} g^{\mu\nu} \partial_\mu \phi \partial_\nu \phi + V(\phi) +\frac{1}{4} g^{\mu\nu} g^{\rho\sigma} F_{\mu\rho} F_{\nu\sigma} + \frac{1}{2} m^2 g^{\mu\nu} A_\mu A_\nu \right) \, .
\eea
In cosmic time $t$ the energy density is then given by,
\bea\label{eq:rhoAx} 
\rho = T_{00} &=& \frac{1}{2} \dot\phi^2 + \frac{1}{2a^2} (\partial_i \phi)^2 + V(\phi) \nn
&+& \frac{1}{2a^2} (\partial_t A_i - \partial_i A_0)^2 + \frac{1}{4a^4} (\partial_i A_j - \partial_j A_i)^2 + \frac{1}{2} m^2 A_0^2 + \frac{1}{2a^2} m^2 A_i^2 \\
&=& \rho_\phi + \rho_A \, , \nonumber
\eea
where $\rho_\phi$ and $\rho_A$ denote the contributions from the inflaton and dark vector respectively.~The pressure density can then be computed as,
\bea
p &=& \frac{1}{3} (g^{\alpha\beta} T_{\alpha\beta} + \rho) 
= \frac{1}{2} \dot\phi^2 - \frac{1}{6a^2} (\partial_i \phi)^2 - V(\phi) \nn
&+& \frac{1}{6a^2} (\partial_t A_i - \partial_i A_0)^2 + \frac{1}{12a^4} (\partial_i A_j - \partial_j A_i)^2 + \frac{1}{2} m^2 A_0^2 - \frac{1}{6a^2} m^2 A_i^2 \\
&=& p_\phi + p_A \, ,\nonumber
\eea
where $p_\phi$ and $p_A$ denote the contributions from the inflaton and dark vector respectively.

In Fourier space, we can write the energy density of the vector field (\eref{rhoAx}) as,
\bea\label{eq:rhoFA0}
\rho_A(\vec x, t) &=& 
\frac{1}{2a^2} \int \frac{d^3 k \ d^3 k'}{(2\pi)^6} e^{i(\vec k + \vec k')\cdot \vec x} 
 \Big\{  [\partial_t \vec A(\vec k,t) \cdot \partial_t \vec A(\vec k',t) ]  + m^2 [\vec A(\vec k,t) \cdot \vec A(\vec k',t)]  \nn
&-& \frac{1}{a^2} \left( [\vec k \cdot \vec k'] [ \vec A(\vec k,t) \cdot \vec A(\vec k',t)] 
- [\vec k \cdot \vec A(\vec k',t)] [\vec k' \cdot \vec A(\vec k,t) ] \right)\\
&-& i A_0(\vec k,t) [\vec k \cdot \partial_t \vec A(\vec k',t) ]
 - [\vec k \cdot \vec k'] [A_0(\vec k,t) A_0(\vec k',t) ] \nn
 &-& i A_0(\vec k',t) [\vec k' \cdot \partial_t \vec A(\vec k,t)]    
+ a^2 m^2 A_0(\vec k,t) A_0(\vec k',t) \Big\} \, . \nonumber
\eea
We can ``bracket'' the expression above and use the definition of the power spectrum \eref{powspclass} to obtain the spatial average $\langle \rho_A(t) \rangle$.~Separating the transverse and longitudinal modes, we have $A_0 = 0$ for the former so last two lines in \eref{rhoFA0} vanish.~After some algebra we have,
\bea \label{eq:rhoTt}
\langle \rho_A^T (t) \rangle &=& \frac{1}{2a^2} \int_0^\infty \frac{dk}{k} \left[ {\cal P}_{\partial_t A_T}(k,t) + \left( \frac{k^2}{a^2} + m^2 \right) {\cal P}_{A_T}(k,t) \right] , \, \nn
\langle \rho_A^T (\tau) \rangle &=& \frac{1}{2a^4} \int_0^\infty \frac{dk}{k} \left[ {\cal P}_{\partial_\tau A_T}(k,\tau) + \left( k^2 + a^2 m^2 \right) {\cal P}_{A_T}(k,\tau) \right] \, .
\eea
in cosmic and conformal time respectively.

For the longitudinal mode, we use~\eref{A0} and its analog in cosmic time,
\bea
A_0(\vec k,t) = \frac{-i \vec k \cdot \partial_t \vec A(\vec k, t)}{k^2 + a^2 m^2} \, ,
\eea
and substitute them into~\eref{rhoFA0}.~After some algebra this gives,
\bea\label{eq:rhoLtau}
\langle \rho_A^L (t) \rangle &=& \frac{1}{2a^2} \int_0^\infty \frac{dk}{k} \left[ \frac{a^2 m^2}{k^2 + a^2 m^2} {\cal P}_{\partial_t A_L}(k,t) + m^2 {\cal P}_{A_L} (k,t) \right] \, , \nn
\langle \rho_A^L (\tau) \rangle &=& \frac{1}{2a^4} \int_0^\infty \frac{dk}{k} \left[ \frac{a^2 m^2}{k^2 + a^2 m^2} {\cal P}_{\partial_\tau A_L}(k,\tau) + a^2 m^2 {\cal P}_{A_L} (k,\tau) \right] \, , 
\eea
in agreement with \cite{Graham:2015rva}.~Note that the expressions in~\eref{rhoTt} and~\eref{rhoLtau} are valid both for the classical and the quantum gauge field upon using the corresponding definitions of the power spectra discussed in the previous section.

\subsection{Analytic study of the energy density spectrum during inflation}\label{sec:appB}

In this section we utilize the (approximate) analytic solutions of the equations of motion to examine the energy density spectrum of the dark vector at the end of inflation.~Starting from~\eref{EOM3} and using $\tau \approx -(a H)^{-1}$ during inflation we define,
\bea
\xi \equiv \frac{\alpha \dot\phi}{2 H f} \, , \qquad \bar m \equiv \frac{m}{H} \, ,
\eea
which allows us to write the equation of motion for the transverse modes as,
\bea
\partial_\tau^2 A_{\pm}(k,\tau) + \left[ k^2 \pm 2\xi \frac{k}{\tau} +  \frac{\bar m^2}{\tau^2} \right] A_{\pm}(k,\tau) = 0 \, .
\eea
Introducing the dimensionless variable,
\bea\label{eq:xdef}
x = - k \tau \approx \frac{k}{aH} \, ,
\eea
the equation of motion then becomes,
\bea \label{eq:EOMx}
\partial_x^2 A_\pm (x) + \left[ 1 \mp \frac{2\xi}{x} + \frac{\bar m^2}{x^2} \right] A_\pm (x) = 0 \, .
\eea
Neglecting the time dependence in $\xi$ and the Hubble parameter, this equation of motion can be solved analytically.~Noting that $x>0$ and using the convention $\xi > 0$, the mode that gets exponentially enhanced is $A_+$ and we can neglect $A_-$ in what follows.~The solution to \eref{EOMx}, once properly normalized, is given in terms of the Whittaker function \cite{Meerburg:2012id},
\bea
A_+(x) = \frac{e^{\pi \xi / 2}}{\sqrt{2k}} W_{-i \xi, \mu} (-2 i x) \, , \qquad \mu = \sqrt{1/4 - \bar m^2} \, .
\eea
From this we can obtain the power spectra as,
\bea
{\cal P}_{\partial_\tau A_+} = \frac{k^3}{2\pi^2} |\partial_\tau A_+|^2 \, , \qquad {\cal P}_{A_+} = \frac{k^3}{2\pi^2} | A_+|^2 \, .
\eea

Starting with the spatially averaged energy density as a function of $\tau$ in~\eref{rhoTt} and the definition of $x$ in~\eref{xdef}, we can write the energy density as,
\begin{align}
\langle \rho_A^T \rangle & = \frac{H^4}{8\pi^2} \int \frac{dx}{x} \ 2k \ x^4 \left[ |\partial_x A_+(x)|^2 + \left( 1+ \frac{\bar m^2}{x^2}  \right) |A_+(x)|^2 \right] \, , \\
\frac{d \langle \rho_A^T \rangle}{d \ln x} & = \frac{H^4}{8\pi^2} 2k \ x^4 \left[ |\partial_x A_+(x)|^2 + \left( 1+ \frac{\bar m^2}{x^2}  \right) |A_+(x)|^2 \right]  \label{eq:drhodlnx} \\
& = \frac{H^4}{8\pi^2} e^{\pi \xi} x^4 \left[ |\partial_x W_{-i \xi,\mu}(-2i x)|^2 + \left( 1+ \frac{\bar m^2}{x^2}  \right) |W_{-i\xi, \mu}(-2ix)|^2 \right] \,.
\end{align}
Note the above is only a function of $x$ under the assumptions that $\tau \approx -(aH)^{-1}$ and $\xi$ is constant, both of which are good approximations during slow-roll inflation.

\begin{figure} [t]
\centering
\includegraphics[width=0.7\textwidth]{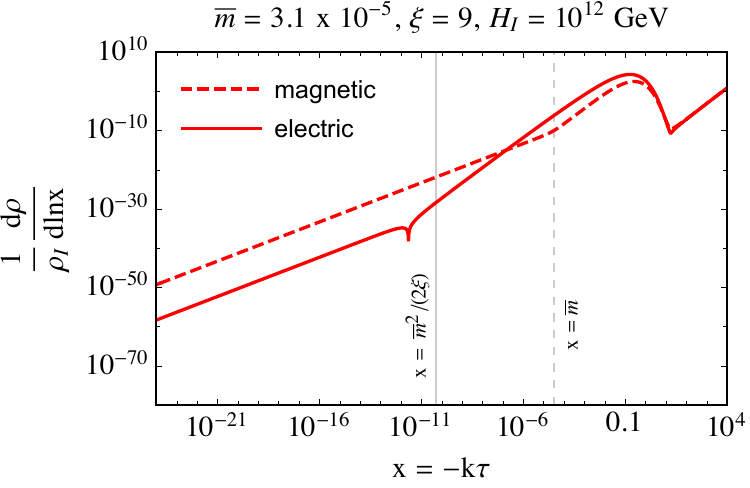}
\caption{Electric (solid) and magnetic (dashed) components of the energy density spectra normalized to the energy density of the inflaton $\rho_I = 3 M_{Pl}^2 H_I^2$. }
\label{Fig:EandB}
\end{figure}

We next separate the contributions to the energy density spectrum into the electric and magnetic components (and drop the brackets for $\rho_A$),
\bea \label{eq:rhoEandB2}
\frac{d\rho_E}{d \ln x} &=&  \frac{H^4}{8\pi^2} e^{\pi \xi} x^4 |\partial_x W_{-i \xi,\mu}(-2i x)|^2 \, ,\nn
\frac{d\rho_B}{d \ln x} &=& \frac{H^4}{8\pi^2} e^{\pi \xi} (x^4 + x^2 \bar m^2)  |W_{-i\xi, \mu}(-2ix)|^2 \, .
\eea
In Fig.~\ref{Fig:EandB} we plot $\frac{d\rho_E}{d \ln x}$ and $\frac{d\rho_B}{d \ln x}$ as a function of $x$ for the listed choice of parameters $\bar m$, $\xi$, and $H$ during inflation, $H_I$.~Both components show a peak at $x \sim 0.1-1$ which indicates the point of maximal tachyonic enhancement.~Moving to lower values of $x$, the magnetic component drops as $x^4$ until it reaches $x = \bar m$ and then decreases as $x^2$.~This behavior can be understood directly from~\eref{rhoEandB2} where for $x < \bar m$ the term $x^2 \bar m^2$ dominates over $x^4$.~This can be traced back easily to $a^2 m^2$ dominating over $k^2$ in the last two terms of \eref{rhoTt}.~The electric component, dominant for $\bar m < x < 0.1$, continues decreasing as $x^4$ until values of $x$ smaller than $\bar m^2 / (2\xi)$ after which point it goes through a kink and then decreases as $x^2$.~It is also clear from the plot that at small $x$ the energy density is dominated by the mass term which we have included in the magnetic component.

Looking more closely at the electric component to understand the kink and the change of slope, the location of the kink occurs where $\partial_x A_+(x)$ changes sign and the electric energy density goes to zero.~It is located at $x < \bar m^2 / (2\xi)$, 
but a compact analytic formula for the exact $x$ is difficult to obtain due to the complicated form of the solution in terms of the Whittaker function.~We can however easily understand the shape to the right and left of the kink by inspection of the equation of motion.~For $\bar m^2 / (2\xi) \ll x < 2\xi$, we can neglect the mass term in \eref{EOMx}.~Then the solution for the tachyonic mode becomes,
\bea \label{eq:masslessol}
A_+(x) = \frac{1}{2k} \sqrt{\frac{4x}{\pi}} e^{\pi \xi} K_1(2 \sqrt{2x \xi}) \, ,
\eea
where $K_1$ is the Bessel function.~We plug this it into~\eref{drhodlnx} and expand for $x\ll 1$ to find,
\bea\label{eq:rhoEap}
\frac{d\rho_E}{d\ln x} = \frac{H^4 \xi e^{2\pi\xi}}{4\pi^3} \left(2\gamma + \ln(2 x \xi) \right)^2 x^4 + {\cal O}(x^{9/2}) \, , \quad  \bar m^2/(2\xi) < x \ll 1 \, ,
\eea
where $\gamma = 0.577$ is the Euler's constant and we see in~\fref{Eapprox} this is an excellent approximation to the full solution in the regime $\bar m^2 / (2\xi) \ll x < 2\xi$.~We also see from the approximate solution in~\eref{rhoEap} that the slope to the right of the kink is $\sim x^4$.
\begin{figure} [t]
\centering
\includegraphics[width=0.7\textwidth]{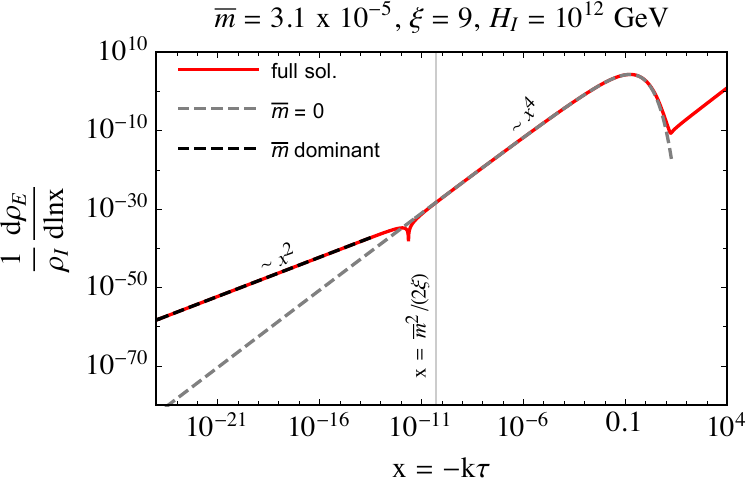}
\caption{Electric component of the energy density spectrum.~For $x > \bar m^2 / (2\xi)$ the full solution (solid red) to the equation of motion is well approximated by the massless one (dashed gray) given in~\eref{masslessol}.~For $x \ll \bar m^2 / (2\xi)$ the mass term dominates and the full solution is approximated by~\eref{massivesol} (dashed black).~The slope changes from $x^4$ on the right of the kink to $x^2$ on the left.}
\label{fig:Eapprox}
\end{figure}

For $x \ll \bar m^2 / (2\xi)$, the mass term dominates in the equation of motion,
\bea
\partial_x^2 A_+(x) + \frac{\bar m^2}{x^2} A_+(x) = 0 \, ,
\eea
which then leads to the approximate solution,
\bea \label{eq:massivesol}
A_+(x) = c_1 x^{\bar m^2} + c_2 x^{1 - \bar m^2} \, , \qquad x \ll \bar m^2 / (2\xi) \, .
\eea
Fixing $c_1$ and $c_2$ to match the normalization of the full solution and plugging into \eref{drhodlnx},
\bea
\frac{d\rho_E}{d\ln x} = \frac{H^4}{8\pi^2} c_1^2 \bar m^4 x^2 + {\cal O}(x^3)\, , \qquad x\ll \bar m^2 / (2\xi) \, , \quad \bar m \ll 1 \, . 
\eea
The solution in this regime is shown in dashed black in~\fref{Eapprox} where we see it is an excellent approximation to the full solution for $x \ll \bar m^2 / (2\xi)$ and the slope goes like $x^2$.

\subsection{Power spectrum of energy density fluctuations}

Here we derive the power spectrum of the energy density fluctuations.~At late times once the dark vector is non-relativistc the energy density is well approximate by the mass term in the lagrangian, $\rho \sim m^2 A^2$.~We can then define the energy density contrast $\delta(\vec x)$ which measures the deviation from the mean energy density,
\bea \label{eq:rhoqu}
\rho(\vec x) = \langle \rho \rangle (1 + \delta(\vec x)) = \frac{1}{2} m^2 \left( \langle \hat{\vec A}(\vec x)^2 \rangle + \hat{\vec A}(\vec x)^2 \right)\, ,
\eea
where we have implicitly dropped the cosmic time variable $t$.~From this we identify,
\begin{align}
\langle \rho \rangle & = \frac{1}{2} m^2  \langle \vec A(\vec x)^2 \rangle , \\
\delta(\vec x) & = \frac{ \vec A(\vec x)^2 }{ \langle \vec A(\vec x)^2 \rangle} = \frac{1}{ \langle \vec A(\vec x)^2 \rangle} 
\int \frac{d^3p}{(2\pi)^3} \frac{d^3q}{(2\pi)^3} e^{i (\vec p + \vec q)\cdot x} \vec A(\vec p) \cdot \vec A(\vec q) \nn
& = \frac{1}{ \langle \vec A(\vec x)^2 \rangle}  \int \frac{d^3k}{(2\pi)^3} e^{i \vec k \cdot \vec x} \int \frac{d^3q}{(2\pi)^3}
\vec A(\vec k - \vec q) \cdot \vec A(\vec q) \\
& \equiv  \int \frac{d^3k}{(2\pi)^3} e^{i \vec k \cdot \vec x} \delta(\vec k) \, ,\nonumber
\end{align}
where we have defined the momentum shift $\vec k = \vec p + \vec q$.~From this we can then read off the Fourier transform of the density contrast,
\bea\label{eq:deltak}
\delta(\vec k) =  \frac{1}{ \langle \vec A(\vec x)^2 \rangle} \int \frac{d^3q}{(2\pi)^3}
\vec A(\vec k - \vec q) \cdot \vec A(\vec q) \, .
\eea

Even though the exponentially enhanced tachyonic modes are highly classical, here we can work explicitly with the quantum field defined in~\eref{Aqu}.~To compute the $\hat{\vec A}(\vec x)^2$ we have, 
\bea
\hat{\vec A}(\vec x)^2 &=& 
\sum_{\lambda, \lambda'} \int \frac{d^3p d^3q}{(2\pi)^6} e^{i(\vec p + \vec q)\cdot x} \vec\epsilon_\lambda(\vec p) \cdot \vec \epsilon_{\lambda'}(\vec q) \nn
&\times& 
\left( A_\lambda(p) a_\lambda(\vec p) + A_\lambda(p)^* a^\dagger_\lambda(-\vec p) \right)
\left( A_{\lambda'}(q) a_{\lambda'}(\vec q) + A_{\lambda'}(q)^* a^\dagger_{\lambda'}(-\vec q) \right) \nn
&=& \int \frac{d^3k}{(2\pi)^3} e^{i \vec k \cdot \vec x}  \sum_{\lambda, \lambda'} \int \frac{d^3q}{(2\pi)^3} \vec\epsilon_\lambda(\vec k - \vec q) \cdot \vec \epsilon_{\lambda'}(\vec q) \\
&\times& 
\left( A_\lambda(k-q) a_\lambda(\vec k - \vec q) + A_\lambda(k-q)^* a^\dagger_\lambda(\vec q - \vec k) \right)
\left( A_{\lambda'}(q) a_{\lambda'}(\vec q) + A_{\lambda'}(q)^* a^\dagger_{\lambda'}(-\vec q) \right) \, .\nonumber
\eea
where again we have defined the momentum shift $\vec k = \vec p + \vec q$.~From~\eref{deltak} we then have,
\bea \label{eq:deltakqu}
\delta(\vec k) &=& 
\frac{1}{\langle \hat{\vec A}(\vec x)^2 \rangle} 
\sum_{\lambda, \lambda'} \int \frac{d^3q}{(2\pi)^3} \vec\epsilon_\lambda(\vec k - \vec q) \cdot \vec \epsilon_{\lambda'}(\vec q) \\
&\times& \left( A_\lambda(k-q) a_\lambda(\vec k - \vec q) + A_\lambda(k-q)^* a^\dagger_\lambda(\vec q - \vec k) \right)
\left( A_{\lambda'}(q) a_{\lambda'}(\vec q) + A_{\lambda'}(q)^* a^\dagger_{\lambda'}(-\vec q) \right) \, .\nonumber
\eea

The space-dependent part of the energy density contrast is obtained from the two-point correlation function,
\bea\label{eq:delta2pt}
\langle \delta(\vec k) \delta(\vec k') \rangle &=& 
\frac{1}{\langle \hat{\vec A}(\vec x)^2 \rangle^2} \sum_{\lambda_1 \lambda_2 \lambda_3 \lambda_4} \int \frac{d^3 q \ d^3 q'}{(2\pi)^6} \vec\epsilon_{\lambda_1}(\vec k - \vec q) \cdot \vec \epsilon_{\lambda_2}(\vec q) \ \vec\epsilon_{\lambda_3}(\vec k' - \vec q^{\ \prime}) \cdot \vec \epsilon_{\lambda_4}(\vec q^{\ \prime}) \nn
&\times& \langle 0|  \left( A_{\lambda_1}(k-q) a_{\lambda_1}(\vec k - \vec q)  \right) \left( A_{\lambda_2}(q) a_{\lambda_2}( \vec q) + A_{\lambda_2}(q)^* a^\dagger_{\lambda_2}(-\vec q) \right) \\
&\times& \left( A_{\lambda_3}(k'-q') a_{\lambda_3}(\vec k' -\vec q^{\ \prime}) + A_{\lambda_3}(k'-q')^* a^\dagger_{\lambda_3}(\vec q^{\ \prime} - \vec k') \right) \left( A_{\lambda_4}(q')^* a^\dagger_{\lambda_4}(-\vec q^{\ \prime}) \right) |0 \rangle \, .\nonumber
\eea
Focusing on the last two lines, there is only one combination of creation and annihilation operators that leads to a space-dependent result,
\bea
\langle 0| a_{\lambda_1} a_{\lambda_2} a^\dagger_{\lambda_3} a^\dagger_{\lambda_4} |0 \rangle &=& (2\pi)^6  
\Big[  \delta_{\lambda_2\lambda_3} \delta_{\lambda_1\lambda_4} \delta^3(\vec k' - \vec q^{\ \prime} + \vec q) \delta^3(-\vec q^{\ \prime} -\vec k + \vec q) \nn
&+&  \delta_{\lambda_1\lambda_3} \delta_{\lambda_2\lambda_4} \delta^3(\vec k - \vec q - \vec q^{\ \prime} + \vec k') \delta^3(\vec q + \vec q^{\ \prime}) \Big] \, .
\eea
Plugging this into~\eref{delta2pt} and keeping only the tachyonic transverse mode $A_+$ we find,
\bea
 \langle \delta(\vec k) \delta(\vec k') \rangle &=& 
 (2\pi)^3 \delta^3(\vec k + \vec k') \frac{1}{\langle \hat{A}_+(\vec x)^2 \rangle^2}  
 \int \frac{d^3 q}{(2\pi)^3} 2 |A_{+}(k-q)|^2 |A_{+}(q)|^2 \\
&=& (2\pi)^3 \delta^3(\vec k + \vec k') \frac{2}{\langle \hat{A}_+(\vec x)^2 \rangle^2} \frac{2\pi}{(2\pi)^3} \int q^2 dqd\cos\theta
\frac{2\pi^2}{(k-q)^3} {\cal P}_{A_+}(k-q) \frac{2\pi^2}{q^3} {\cal P}_{A_+}(q) \nn
&=&  (2\pi)^3 \delta^3(\vec k + \vec k') \frac{2 \pi^2}{\langle \hat{A}_+(\vec x)^2 \rangle^2}
\int dq \ dp \ q^2 \frac{p}{kq} \frac{1}{p^3 q^3} {\cal P}_{A_+}(p) {\cal P}_{A_+}(q) \, ,\nonumber
\eea
where we have used~\eref{powquantum} and the change of variables from $q, \cos\theta$ to $q,p$ using,
\bea
\vec p = \vec k - \vec q\, , 
~~p^2 = k^2 + q^2 - 2 kq \cos\theta\, , 
~~d\cos\theta = - \frac{p}{kq} dp\, , 
~~dq d\cos\theta \rightarrow \frac{p}{kq} dq dp \, ,
\eea
as well as trivially performed the $d\phi$ integral.~Defining the power spectrum for the energy density contrast in the quantum case (and for transverse modes) as,
\bea \label{eq:powdqu}
\langle \delta(\vec k) \delta(\vec k') \rangle &=& 
(2\pi)^3 \delta^3(\vec k + \vec k') \frac{2\pi^2}{k^3} {\cal P}_{\delta}(k) \, ,
\eea
we arrive finally at the expression given in~\eref{deltaPS},
\bea \label{eq:powdeltaqu}
{\cal P}_{\delta}(k) &=& \frac{k^2}{\langle \hat{A}_+(\vec x)^2 \rangle^2}
 \int_{|q-k|<p<q+k} dq \ dp \ \frac{1}{q^2 p^2}  {\cal P}_{A_+}(p) {\cal P}_{A_+}(q) \, , \nn
&&\langle \hat{A}_+(\vec x)^2 \rangle^2 = \left[ \int_0^\infty \frac{dk}{k} {\cal P}_{A_+} (k)  \right]^2 \, .
\eea
This is the power spectrum of the energy density contrast for $\rho \sim m^2 A^2$.~For completeness we have also computed the power spectrum for the energy density contrast corresponding to the kinetic term in the energy density for which we find the same expression as~\eref{powdeltaqu} with $A_+ \to \partial_t A_+$.~For the longitudinal mode we find the same result as in~\cite{Graham:2015rva}.

\subsection{Numerical procedure for solving equations of motion}

Here we sketch the numerical solutions to the equations of motion and how the input spectra are obtained.~We have integrated the equations of motion for the longitudinal and transverse vector perturbations at linear order in cosmic time\footnote{We have used the~\href{https://computing.llnl.gov/projects/sundial}{SUNDIALS package}, “SUite of Nonlinear and DIfferential/ALgebraic equation Solvers” for the numerical integration.},
\bea
\ddot A_L(\vec k,t) + \frac{3k^2 + a^2(t) m^2}{k^2 + a^2(t) m^2} H \dot A_L(\vec k,t) + \left( \frac{k^2}{a^2(t)} + m^2 \right) A_L(\vec k,t) &=& 0 \,, \\
\ddot A_{\pm}(k,t) + H \dot A_{\pm}(k,t) + \left[ \frac{k^2}{a^2(t)} \mp \frac{\alpha}{f} \dot\phi \frac{k}{a(t)} + m^2 \right] A_{\pm}(k,t) &=& 0 \, ,
\eea
together with the background inflaton equation of motion,
\be
\ddot \phi +3 H \dot \phi + V^\prime=0 \,, 
\ee
where $V^\prime = d V(\phi)/d \phi$.~Working in the regime of no dark vector backreaction during single-field inflation, the Hubble parameter is given by,
\be
H^2 = \frac{1}{3 M_{PL}^2} \left( \frac{\dot \phi^2}{2} + V(\phi) \right) \,, 
\ee
and the number of e-folds is obtained by integrating,
\be
\frac{d \ln a}{dt} = H \,.
\ee
We start the integration in the slow-roll regime with initial field velocity,
\be
\dot \phi = \frac{V^\prime}{3H} \,,
\ee
and the initial value of the inflaton field for any given inflationary potential $V(\phi)$ will set the total number of e-folds $N_e$ up to the end of inflation at $\epsilon_{end}=\dot \phi^2/(2H^2 M_{PL}^2) = 1$.~For example, for the models considered in~\fref{energyspecsteep}, $\phi(0)=24 M_{PL}$ gives $N_e\simeq 72$ for a chaotic quartic potential, while $\phi(0)=17 M_{PL}$ gives $N_e\simeq 73$ for a chaotic quadratic potential, $\phi(0)= M_{PL}$ gives $N_e\simeq 73$ for the hilltop quadratic potential with $v=6 M_{PL}$, and $\phi(0)=60 M_{PL}$ gives $N_e\simeq 62$ for the axion-like model with $\Lambda=24 M_{PL}$.

For each $k$ mode, we start the integration with the vector fluctuations initially in the Bunch-Davies vacuum defined as,
\begin{align}
A_{\lambda}^{(R)}(k,0) &= \frac{1}{\sqrt{2 \omega_k}}\,,
&A_{\lambda}^{(I)}(k,0) &= 0 \,, \\
\dot A_{\lambda}^{(R)}(k,0) &= 0 \,,
&\dot A_{\lambda}^{(I)}(k,0) &= -\frac{\omega_k}{\sqrt{2}}\,,
\end{align}
where $\lambda$ refers to either longitudinal or transverse mode, $\omega_k^2 = k^2/a_0^2 + m^2$,  $a_0$ is the initial value of the scale factor, and  $R$($I$) refers to the real (imaginary) component of the perturbation. Therefore, the initial vacuum power spectrum is given by,
\bea
{\cal P}_{A_\lambda}(k,0) &=& \frac{k^3}{2 \pi^2 a_0^3} |A_\lambda(k,0) |^2 = \frac{k^3}{4 \pi^2 \omega_k a_0^3} \,, \\
{\cal P}_{\dot A_\lambda}(k,0) &=& \frac{k^3}{2 \pi^2 a_0^3} | \dot A_\lambda(k,0) |^2 = \frac{k^3 \omega_k}{4 \pi^2 a_0^3 } \,.
\eea     
Two-point functions and energy densities for the dark vector perturbations obtained directly from the momentum integration of the power spectrum are clearly UV divergent and must be regularized.~In an expanding universe this can be accomplished using the adiabatic regularization method~\cite{Parker:2009uva} which is based on a WKB-type expansion in powers of the time derivatives of the scale factor and frequency modes.~For our purposes, given that we are interested in particle production effects driven by the tachyonic instability, it is enough to regularize the expressions at zero-order, i.e.\,we only subtract the vacuum contribution as in Minkowsky space, 
\bea
{\cal P}^{reg}_{A_\lambda}(k,t) &=& {\cal P}_{A_\lambda}(k,t) - {\cal P}_{A_\lambda}(k,0) \,, \\
{\cal P}^{reg}_{\dot A_\lambda}(k,t) &=& {\cal P}_{\dot A_\lambda}(k,t) - {\cal P}_{\dot A_\lambda}(k,0) \,.
\eea     
On the left hand side in~\fref{energyspec} the dotted line shows the subtracted vacuum contribution.~We see all the spectra are above this line up to modes $k/a_{end} \sim O(10) \Hend$ showing the particle production effect.~Higher momentum modes stay in the Bunch-Davies vacuum and the partial subtraction we have performed leads to a spectra below the vacuum one.~This signals the absence of the tachyonic instability and particle production effects for these modes.    


\bibliographystyle{JHEP}
\bibliography{ClumpyDMrefs}

\end{document}